\documentstyle[12pt]{article}
\begin{document}


\bigskip {\it}

\vskip  15cm

\centerline{\bf GEOMETRICAL  DESCRIPTION  OF  VORTICES }
  
  \vskip 1cm 
  
       \centerline{\bf  IN GINZBURG-LANDAU BILLIARDS.}

 \vskip 2cm 
\centerline{ E. Akkermans (1) and K. Mallick (2)}
\vskip 1cm

\centerline{(1) Lab. de Physique des Solides, LPTMS, 91405, Orsay Cedex, France}

\centerline{and Physics Dept. Technion, Israel Institute of Technology, Haifa 32000, Israel.}

\vskip 1cm 
\centerline{ (2) Physics Dept. Technion, Israel Institute of Technology, Haifa 32000, Israel}

\centerline{and Service de Physique Th\'eorique, Centre d'\'etudes Nucl\'eaires de Saclay,} 

\centerline{91191, Gif sur Yvette cedex, France.} 
\vskip  10cm 

Les  Houches Summer School (session LXIX),
 Topological aspects of low dimensional
 systems, July 1998


\eject
\centerline{\bf CONTENTS}

{\bf 1. Introduction}

{\bf 2.  Differential manifolds}

2.1 Manifolds

2.2 Differential forms and their integration 

{\it 2.2.1 Tangent space}

{\it 2.2.2 Forms}

{\it 2.2.3 Wedge-Product}

{\it 2.2.4 The exterior derivative}

{\it 2.2.5 Closed and exact forms}

{\it 2.2.6 Integration of forms}

{\it 2.2.7 Theorem of Stokes}

 2.3 Topological invariants of a manifold

 {\it  2.3.1 Motivations }

 {\it  2.3.2 The Euler-Poincar\'e Characteristic }

 {\it  2.3.3 De Rham's cohomology }

 2.4  Riemannian Manifolds and Absolute Differential Calculus

 {\it 2.4.1 Riemannian Manifolds }

  {\it 2.4.2 Covariant derivative. Connexion and curvature form }

 2.5 *The Laplacian

 2.6 Bibliography

{\bf 3. Fiber bundles and their topology}

3.1  Introduction

3.2  Local symmetries. Connexion and curvature

3.3  Chern  classes

3.4  Manifolds with a boundary: Chern-Simons classes

{\it 3.4.1 The Gauss-Bonnet theorem}

{\it 3.4.2 Surfaces with boundary}

{\it 3.4.3  Secondary characteristic classes}

3.5  *The Weitzenb\"ock formula 

{\bf 4. The dual point of Ginzburg-Landau equations for an infinite system}

4.1 The Ginzburg-Landau equations 

4.2 The Bogomol'nyi identities

{\bf 5. The superconducting billiard}

5.1 The zero current line

5.2 A selection mechanism and topological phase transitions

5.3 A geometrical expression of the Gibbs potential for finite systems

\eject

\section{ Introduction}

\vskip 0.3cm

In these notes we discuss the topological
 nature of some problems in condensed 
matter physics. This topic has been widely studied
  in various contexts. 
In statistical mechanics, the possible stable defects
 in an ordered system   have been classified
 according to the  nature of the order 
parameter   (e.g. scalar, vector, matrix)
 and the space dimensionality of the system using
 homotopy groups \cite{toulouse}.
 Then, the discovery of the quantum Hall effects and 
the role played by stable integers or
 rational numbers for systems with few or no conserved 
quantum symmetries have motivated several
 topological models of quantum
  condensed matter systems  
 \cite{fradkin,avron,thouless} .
A combination of these two ideas of
  defects classification and microscopic quantum models 
has been used in the description of superfluid ${^3}He$ \cite{volovik}.

Here, instead of trying an exhaustive  review 
of problems where topological ideas may play a role, we  present 
  the basic constituents needed in a  geometrical 
description and   calculate the related topological numbers.
The ideas and methods developed in mathematics and mathematical
 physics to solve problems 
in  geometry and topology are pretty sophisticated 
and sometimes expressed in a way 
unfamiliar to the physicist. 
We  adopt the language of differential geometry
 to present this subject, since
  it is  adapted to develop some 
intuition towards more elaborate concepts like fiber bundles,
 connexions and 
  topological invariants.

 In the last two sections,
 we shall discuss the problem of superconducting billiards
 within the Ginzburg-Landau approximation. This 
problem is interesting for several reasons. First, it is
 a non trivial example on which  
topological methods 
naturally apply to give an elegant solution
 to the calculation of the ground 
state energy. It is also a situation for which
 one  can address the question of transition between 
different values of a topological number
 controlled  by the boundary of the system. This 
 question is similar to the transition between
 quantum Hall plateaus.

  Such problems  are not only of academic interest. Our motivation 
  has been triggered  by a set of new experimental results 
  obtained on small size aluminium disks
 \cite{geim,deo} in a regime where their 
  radius $R$ is comparable with both the coherence length  $\xi$
 and the London penetration
 depth $\lambda.$ The magnetization, 
  as a function of the
 applied magnetic field,  presents  a series of jumps
 with an overall shape  reminiscent of type-II superconductors,
 although  a macroscopic sample of aluminium is 
 a genuine type-I superconductor. These notes end with
 a theoretical analysis of these experimental results.

\section{Differentiable Manifolds}


 \subsection{ Manifolds}
\vskip 0.3cm

    A   differentiable manifold $M$
 of dimension $n$ is a
 space which looks  locally  
like an open set of ${\bf R}^n$.
  On  the vicinity of each point of $M$
  one can define a {\bf  local coordinate system} where
 each point is represented   by a set of 
$n$ real numbers $(x^1,...,x^n)$.
 There exists 
geometrical spaces  which are not differentiable manifolds like  
arithmetical ensembles  (${\bf Q},{\bf  Z}/5{\bf Z}$), fractals,
 objects with branching points 
(Feynman diagrams). On the other hand vector spaces, spheres,
  projective spaces,
 matrix  groups $GL(n,{\bf R}), SO(2), SO(3)$ (direct isometries),
 $SU(2), SU(n)$ (complex 
isometries) are differentiable manifolds. 
In order to do differential calculus on a manifold,
 different coordinate systems are required to be {\bf compatible} 
 with each other: the local  transformations between
 one system of coordinates to another have to be smooth and invertible.
 A   quantity attached
 to a manifold is said to be {\bf  geometrical} 
(or {\it  intrinsic}) if it is independent of
 the choice of a coordinate system.
 \hfill\break
 {\it Example:}  the sphere $S^2$ in Euclidian
  space ${\bf R}^3$ can be endowed
 with local cartesian coordinates, or with spherical coordinates
 (latitude and longitude). None of these two systems is global, for
 instance the spherical coordinates are singular at the poles,
 although from a  geometrical point of view all points on the sphere
 are equivalent. Therefore the special role played by the poles
 are an artifact   of the spherical  coordinate system. On the other hand
 the tangent plane passing through a point
 of the sphere is an intrinsic object,
 although its equation looks very different in cartesian and spherical
 coordinates.

\vskip 0.3cm
\subsection{Differential forms and their integration}

 \subsubsection{Tangent space}

  Consider a manifold M and  $p$ one of its points.
  Let $\gamma$ be a  curve in M passing through $p$.
 In a local coordinate system, the curve
 $\gamma$ is given by 
 \begin{eqnarray}
   \gamma :  [-1, 1] &\to& M \nonumber \\
              t &\mapsto&  (x^1(t),...,x^n(t))
\label{curve}
 \end{eqnarray}
 For $t=0$,  $\gamma$ passes through the point  $p$ represented
 by the coordinates $ (x^1(0),...,x^n(0))$.
 A {\bf tangent vector} 
 to the manifold $M$ at the point $p$
 is by definition a tangent vector to a curve passing
 through $p$. For instance, the curve  $\gamma$
  defines  a tangent vector at $p$
${\vec v} \left( = \frac {d\gamma}{dt}\right)_{t=0}$  
 with components:
 \begin{equation}
   v^i = \left( \frac{ d x^i} {d t} \right)_{t=0} \,\,\,.
 \end{equation}
  If one uses a different set of local  coordinates
 $(X^1,..., X^n)$  in the vicinity of $p$, one finds  that
 the  components of ${\vec v}$ are given by
 \begin{equation}
   V^i = \left( \frac{ d X^i} {d t} \right)_{t=0} \,\,\,.
 \end{equation}
 The {\bf transformation law} from one set of components
 to the other is
\begin{equation}
     V^i = \left( \frac{\partial X^i}{\partial x^j} \right)_{p} v^j \,\,\,\, .
\label{trans}
\end{equation}
 We  use  Einstein's  convention of summation upon repeated indexes.
 The  transformation law of the components
 of ${\vec v}$ is inverse of that of the partial derivatives
 $ \frac{\partial}{\partial x^j}$. Hence in `classical' (XIX century)
 mathematical literature, a tangent vector was said to be
 {\it contravariant} and  the transformation  formula
 was used as a {\it definition}:  a tangent vector is an object
 whose components transform according to (\ref{trans}) under
 a  change of the local coordinate system.

  The set   of all  the  tangent vectors at $p$
  to all curves  $\gamma$ included in $M$ and passing
through $p$  is a vector space, called the
 {\bf the tangent space} of $M$ at $p$ and denoted by ${T_p}M$.
 It has the same dimension as the manifold itself.
  With the help of (\ref{trans}),   one can define
   a  tangent vector, 
 as an invariant object  that does not depend on the coordinate
 system:
 \begin{equation}
 {\vec  v} =   v^i \left( \frac{\partial}{\partial x^i} \right)_{p} = 
              V^i  \left(  \frac{\partial}{\partial X^i}  \right)_{p}
\end{equation}
 The  vectors
\begin{equation}
 e_i =\left( \frac{\partial}{\partial x^i}\right)_{p}
\,\,\,\,\,  i=1,...,n 
\label{base}
\end{equation}
  are   a basis of the tangent space ${T_p}M$.
 This basis is {\it  local} and depends  on the point $p$.
 Usually one does not write explicitly the dependence  on $p$.
 However it is extremely important to keep in mind
 that two tangent spaces  ${T_p}M$ and  ${T_{p'}}M$ at
 two different points $p$ and $p'$ are {\it different}
 vector spaces. It is not possible a priori  to compare 
 (i.e add, subtract) two tangent  vectors ${\vec v}$ 
 and ${\vec {v'}}$  at  $p$ and $p'$. A simple way to understand
 that is to realise that the transformation law (\ref{trans})
  is point dependent.  Therefore ${\vec v}$ 
 and ${\vec {v'}}$ could by chance have the same components
 in a special  coordinate system but  different components in
 another system. So it would be a mistake to say  that
 they are `equal'.

  The {\it total} {\bf tangent space } $TM$  of $M$ is the collection
 of all tangent vectors at any point of $M$. Hence
 $TM = \bigcup_p  T_{p}M$. There  is a natural function
 $\pi$ from $TM$ to $M$ called {\it projection: } to each
 ${\vec v}\in TM$ is associated   $p=\pi({\vec v})$ which is  the point of
 $M$ at which  ${\vec v}$ is tangent to $M$. It can be  shown that $TM$
 is a manifold of dimension $2n$: if ${\vec v}$
  belongs to $TM$ one needs $n$ components to
 specify at which point $p$ the vector  ${\vec v}$ is tangent
 and $n$ more components to specify ${\vec v}$ in $T_{p}M$.

 By definition a {\bf vector-field} is a smooth field of tangent
 vectors to  the manifold $M$, i.e.  to each point $p$ is associated 
 a vector ${\vec F(p)}$ tangent to $M$ at $p$.
 In  a local coordinate system  a vector-field
  is expressed as
\begin{equation}
{\vec F}(x^1,...,x^n) = F^i(x^1,...,x^n) \frac{\partial}{\partial x^i}
\end{equation}

\subsubsection{Forms}

  A {\bf 0-form } is a smooth scalar valued function on the
 manifold $M$.
  A {\bf 1-form }  is a linear function on vectors.
 Consider first the case of 
 ${\bf R}^3$.
 A vector ${\vec v}= (x,y,z)$ in ${\bf R}^3$ is written  in a  basis 
${\vec v}= x {\vec i} + y {\vec j} +z {\vec k}$. Let $\omega$
 be a 1-form on ${\bf R}^3$.
  By linearity, $\omega ({\vec v})$
  is given by
\begin{equation}
\omega ({\vec v}) = x \omega ({\vec i}) +
 y \omega ({\vec j}) +z \omega ({\vec k}).
\end{equation}
 Thus the question 
  of calculating  $\omega({\vec v})$ boils down to the question
 of evaluating it on the  vectors of the basis 
 ${\vec i}, {\vec j}, {\vec k}$.
 In  ${\bf R}^3$,  we define the 1-form $dx$ by $dx ({\vec i}) =1, 
 dx({\vec j}) = 0 $ and $dx ({\vec k}) = 0 $ and
 equivalently for the 1-forms $dy$ and $dz$.
 We have shown that
$$ \omega = \omega ({\vec i}) dx + \omega ({\vec j}) dy +
 \omega ({\vec k}) dz \,\,\, .$$
 The triplet $(dx,dy,dz)$ is a basis for 1- forms
 called the {\it dual-basis} of  $({\vec i}, {\vec j}, {\vec k})$.
 \hfill\break
 More generally, one can consider a field
 of 1-forms on  ${\bf R}^3$  i.e.   a 1-form $\omega$ with coefficients
 that vary from point to point. It  can be written as 
 $$\omega = A(x,y,z) dx + B(x,y,z) dy + 
  C(x,y,z) dz \,\,\,\,   .$$
  The action of  $\omega$  on  a vector field ${\vec F} = f(x,y,z)
 {\vec i} + g(x,y,z) {\vec j} + h(x,y,z) {\vec k}$ is 
\begin{equation}
\omega({\vec F})(x,y,z) = Af+Bg +Ch.
\end{equation}
 A {\bf k-form} 
 is a smooth multilinear and antisymmetric
 function on k-tuples of tangent vectors to M, all 
 of them  tangent at the same point.
This can be implemented by considering two vectors in the plane, ${\vec v_1} =
x_1 {\vec i} + y_1 {\vec j} $ and ${\vec v_2} =
x_2 {\vec i} + y_2 {\vec j} $. Let $\phi$ be a 2-form, then,
\begin{eqnarray}
\phi ({\vec v_1},{\vec v_2}) &=& x_1y_1 \phi ({\vec i},{\vec i}) +
 x_1y_2 \phi ({\vec i},{\vec j}) + 
 x_2y_1 \phi ({\vec j},{\vec i}) + x_2y_2 \phi ({\vec j},{\vec j})
\\
\nonumber
&=&( x_1y_2 - x_2y_1) \phi({\vec i},{\vec j}) 
\end{eqnarray}
   because the antisymmetry condition implies $ \phi ({\vec i},{\vec i})
 = \phi ({\vec j},{\vec j}) = 0$ and $ \phi ({\vec i},{\vec j})
 = - \phi ({\vec j},{\vec i})$.
The requirement that differential forms be
 antisymmetric  comes from the fact that we need to keep track 
of orientation.

\hfill\break
{\it  Examples: } \hfill\break
 1. In ${\bf R}^3$ the 2-form $dx \wedge dy$ is defined by
 $dx \wedge dy({\vec i},{\vec j}) =1 ,
 dx \wedge dy({\vec j},{\vec k}) =0$
 and $dx \wedge dy({\vec k},{\vec i}) =0$ and equivalently
 for the 2-forms $dy \wedge dz$ and
 $dz \wedge dx$.\hfill\break
 2. More generally,  in  ${\bf R}^n$, the dual basis of the canonical
 basis $(e_1,...,e_n)$ is denoted by $(dx^1,...,dx^n)$.
 It satisfies $dx^i(e_j) = \delta^i_j$, where $\delta^i_j$
  is the Kronecker delta.
 One defines
 the $k$-form
 $dx^{i_1} \wedge dx^{i_2} \wedge ...\wedge  dx^{i_k}$, such that
if $({\vec v_1},{\vec v_2},...,{\vec v_k})$ 
 is a $k$-tuple of vectors of  ${\bf R}^n$,
 the quantity
  $dx^{i_1} \wedge dx^{i_2} \wedge ...\wedge  dx^{i_k} 
({\vec v_1},{\vec v_2},...,{\vec v_k})$
 is the $k\times k$ determinant of the components of the  vectors
 $ {\vec v_i}$  along the directions defined by 
  $(e_{i_1}, e_{i_2}...e_{i_k})$. 
  The geometric interpretation
  of this number  is known: it is
 the volume of the projection of
 the parallelepiped generated by $({\vec v_1},{\vec v_2},...,{\vec v_k})$ 
 on  the linear space spanned  by  $(e_{i_1}, e_{i_2},...,e_{i_k})$.

 From example 2, one can prove that the set of $k$-forms
 $(dx^{i_1} \wedge dx^{i_2} \wedge ...\wedge  dx^{i_k})$
 with $i_1 < i_2<...<i_k$ is a basis of the vector space
 of the   $k$-forms in  ${\bf R}^n$.  Hence the
 space of $k$-forms has dimension ${n !} \over {k! (n-k )!}$.
 In particular all $n$-forms on  ${\bf R}^n$ are proportional
 to $(dx^{1} \wedge dx^{2} \wedge ...\wedge  dx^{n})$,
 which is nothing but the determinant.

\hfill\break
 On an general  manifold $M$, one constructs $k$-forms
 locally for each tangent space ${T_p}M$.
  For instance, a {\bf 1-form }  
 $\omega$ on M is  a smooth linear  function on ${T_p}M$
 for every point $p$ on M. Similarly,
 a  $k$-form $w$ on a manifold $M$ is  a smooth
 collection of $k$-forms on each tangent space ${T_p}M$.
 For each $p$ one defines  a local basis 
 $((dx^1)_p,...,(dx^n)_p)$ dual to the basis (\ref{base})
 $(e_1=\left( \frac{\partial}{\partial x^1}\right)_{p}, ...,
  e_n=\left( \frac{\partial}{\partial x^n}\right)_{p})$
 of ${T_p}M$. Then any $k$-form $w$ on  $M$ can 
 be written in a local system of coordinates as follows:
\begin{equation}
 w = \sum A_{i_1,..., i_k} dx^{i_1} \wedge dx^{i_2} \wedge ...\wedge 
 dx^{i_k}
 \label{kform}
\end{equation}
 where the $A_{i_1,..., i_k}(p)$ are real functions. Once again,
 the $p$ dependence is not explicitly  stated.

\subsubsection{Wedge-Product}

 A 2-form $\psi$ can be constructed by forming the
  {\it  wedge-product}
  $\psi = \omega_1 \wedge \omega_2$ 
of two smooth 
1-forms $\omega_1$ and $\omega_2$ via:
\begin{equation}
\psi ({\vec u}, {\vec v}) =
 \omega_1({\vec u}) \omega_2 ({\vec v}) -
 \omega_1({\vec v}) \omega_2 ({\vec u})
\end{equation}
for any vectors $\vec u$ and $\vec v$.
 More generally, if $\phi$ is a $k$-form and $\psi$ a $l$-form
 one can construct a $(k+l)$-form $ \phi \wedge \psi$
 which acts on $(k+l)$-tuples of tangent vectors at a point,
 by antisymmetrizing correctly the product  $ \phi \wedge \psi$.
 We  shall not write the explicit formula here. It the same as the
 one used to construct a fermionic (antisymmetric) wave-function
 of  $(k+l)$  variables starting with two fermionic wave functions
 of respectively $k$  and $l$  variables. One has in particular
\begin{equation}
 \phi  \wedge \psi = (-1)^{kl}  \psi  \wedge \phi \,\,\,\, .
\label{sign}
\end{equation}

 \subsubsection{The exterior derivative}

 The exterior
 derivative  $d$  generalizes the usual
 operations of vector calculus.
  The exterior derivative  $d$  
 transforms 
  $k$-forms into $(k+1)$-forms.
 Let for 
instance A be a 0-form (a function), its exterior derivative
 $dA$ is the 1-form
\begin{equation}
dA = \frac {\partial A}{\partial x^1} dx^1 + ... + 
\frac {\partial A}{\partial x^n}dx^n  \,\,\, .
\label{0-form}
\end{equation}
 If ${\vec v}$  is a tangent vector then 
$$ dA ({\vec v}) = \frac {\partial A}{\partial x^i} v^i $$
is the rate of variation of the function $A$ in the direction
 of  ${\vec v}$. This quantity is usually denoted by
 $(\nabla{A}).{\vec v}$ or $\nabla_{\vec v}{A}$.
 We have seen  that a  $k$-form $w$ can be written in a local basis
 $$ w = \sum A_{i_1,..., i_k} dx^{i_1} \wedge dx^{i_2} \wedge ...\wedge 
 dx^{i_k}$$
 where the $A_{i_1,..., i_k}$ are real functions.
 The exterior derivative  operates  on $w$ by
 acting on each of the coefficients  $A_{i_1,..., i_k}$
 via (\ref{0-form}).
  For example, for a 1-form $\phi = A dx + B dy +Cdz$, 
 one obtains
\begin{eqnarray}
d \phi &=& dA \wedge  dx + dB \wedge dy + dC\wedge  dz \nonumber \\
    &=&  (C_y - B_z) dy \wedge dz + (A_z - C_x) dz \wedge dx 
 + (B_x - A_y) dx \wedge dy
\end{eqnarray}
An important property of the exterior derivative
  is that it gives 0 when applied twice, 
\begin{equation}
d^2 \phi =0
\end{equation}
 for any $k$-form $\phi$. This follows from the Schwartz identity
 for partial derivatives:
 $$ \frac  {\partial^2 A} {\partial x_i \partial x_j}
 = \frac  {\partial^2 A} {\partial x_j \partial x_i} $$
 There is a Leibniz rule for the exterior derivative of
 a $k$-form  $\phi$ and a $ l$-form $\psi$
\begin{equation}
d(\phi \wedge \psi) = (d\phi) \wedge \psi + (-1{)^k} \phi \wedge d \psi
\end{equation}

 \hfill\break
{ \it Examples: }
In the Euclidean space ${\bf R}^3$, a vector field 
$\vec F$ is defined at each point $(x,y,z)$ by  
${\vec F}= A {\vec i} + B {\vec j} + C {\vec k}$,
 where $A,B,C$ are smooth functions 
of the coordinates. And 
  let    $f$ be a function. Its gradient is the 
vector field $\nabla{f} = f_x {\vec i} + f_y {\vec j} + f_z {\vec k}$.
 Similarly,  the application 
of the exterior derivative $d$ on the  0-form (function)
 $f$ gives the 1-form 
$df = f_x dx + f_y dy + f_z dz\,\,\, .$ Hence,
 to the 1-form $df$ is associated the vector field 
$\nabla  f$. The action of the rotational
 on the vector field ${\vec F} $ gives another 
vector field
$$\nabla\times {\vec F} = 
(C_y - B_z) {\vec i} + ( A_z - C_x ) {\vec j} +
 (B_x - A_y) {\vec k} \,\,\,\,\, ,$$
 and the  exterior derivative $d$ operates
 on a 1-form $ \phi = A dx + B dy +C dz$ to give the 
 corresponding  2-form 
$$d \phi = (C_y - B_z) dy \wedge dz +
 ( A_z - C_x ) dz \wedge dx + (B_x - A_y)dx \wedge dy \,\,\,\, .$$
Finally,  one can check that
the divergence of a vector field corresponds to
  $d$ acting on a 2-form to  generate  a 3-form.
 In summary, gradient, curl and divergence result from the 
application of $d$ to 0-forms, 1-forms and 2-forms respectively.
The relations $\nabla\times\nabla{A} = {\vec 0}$ 
and $\nabla.\nabla\times F =0$ are simply  a consequence of $d^2 = 0$.

\vskip 0.3cm
 \subsubsection{Closed and exact forms.}
\vskip 0.3cm

If $\phi$ is a differential form defined on a manifold
 $M$ with the property  $ d \phi =0$, then 
$\phi$ is said to be {\bf  closed}.
 If it has the property that $\phi = d \psi$ for some
 form $\psi$ on each point in 
$M$, then $\phi$ is  {\bf exact}.
 It follows from these definitions and from  $d^2 =0$, 
that {\bf  every exact form is closed.} 
 But the reciprocal is not true with the 
important exception that on a simply connected domain
 $M$, i.e.  a domain in which  every closed curve 
can be continuously deformed to a point through deformations 
that remain  in $M$, every closed 1-form is exact.
 We shall discuss later in more detail the 
1-form 
\begin{equation}
\omega = {{-y} \over {x^2 + y^2}} dx + {{x} \over {x^2 + y^2}}dy.
\label{angle}
\end{equation}
It is closed $(d\omega = 0)$
 but not exact on $M = {\bf R}^2 \backslash (0,0)$.
\vskip 0.3cm

 \subsubsection{Integration of forms.}
\vskip 0.3cm

A $k$-dimensional submanifold (or $k$-chain) of a manifold
 $M$  of dimension 
$n \geq k$ is a subset which can be parametrized with only
 $k$ coordinates.
 In this terminology, a 0-chain is a point, a 1-chain is a curve
  and a 2-chain a surface.
$k$-forms are the right objects to be integrated over  $k$-chains.
 Indeed  $k$-forms were invented by Elie Cartan for this purpose!
 (Their similarity with determinants is not by chance: when  changing
 variables in an oriented integral
 one must   multiply the integrand by  a  determinant  and 
 $k$-forms were built to  produce such a determinant by a 
  change of  the  local coordinate system.)
 We shall simply consider 
 the integral of 1-forms over 1-chains
 in the Euclidean space ${\bf R}^3$.
Let ${\cal C}$ be an oriented smooth curve (i.e. a 1-chain)
 parameterized by ${\vec r}(t) = (x(t), y(t), z(t))$ for
 $t \in I = [-1,1]$ and let
 $\omega = A dx + B dy + C dz$ 
 be a smooth 1-form on ${\cal C}.$  Then,
\begin{equation}
\int_{{\cal C}} \omega = \int_{-1}^{1}
\left( A(x,y,z) x'(t)  + B(x,y,z) y'(t)  + C(x,y,z) z'(t) \right) dt
\end{equation}
Let  ${\vec F}= A {\vec i} + B {\vec j} + C {\vec k}$
 be  the vector field which corresponds  to the 1-form 
$\omega$. We obtain the more familiar expression
\begin{equation}
\int_{{\cal C}} \omega = \int_{{\cal C}} {\vec F}.{\vec dr}
\end{equation}
 which represents  the work of ${\vec F}$ 
 along the curve ${\cal C}$.

 \hfill\break
{\underline {Exercise}: The winding number \hfill\break
  1. Show that on the unit circle
 $S^1$ parameterized by $r( \theta) = (\cos \theta,\sin \theta)$
 in the plane, the 1-form $\omega$ given by (\ref{angle})
 is such that $\int_{S^1} \omega = 2 \pi$.
\hfill\break
 2.  Show that if $\gamma$ is a path connecting two points
 $P$ and $Q$ of the plane ${\bf R}^2 \backslash (0,0)$
 the integral $\int_{\gamma}\omega$ mesures the difference
 of the polar angles of $Q$ and $P$, the center of the polar
 coordinates being (0,0). \hfill\break
 3. Deduce from 2. that if  $\gamma$  is a closed path
 that  encircles $n$ times
 the point (0,0),
 $ W(\gamma,(0,0)) \equiv  \frac{1}{2\pi} \int_{\gamma}\omega = n$.
  In particular if  $\gamma$  is a closed path that
 does not  encircle (0,0), this integral is equal to zero. \hfill\break
 4. The mapping $ W(\gamma,(0,0))$
 defined from the space of closed curves
 in  ${\bf R}^2 \backslash (0,0)$ to the set of rational
 integers  ${\bf Z}$ is called {\bf the winding number}.
  It allows to classify different type of curves
  in ${\bf R}^2 \backslash (0,0)$. It is a simple example
 of a {\bf topological invariant} (see section 2.3).
 Two curves with the same winding number are said
 to be {\it  homologous.}

 \subsubsection{Theorem of Stokes:}

Let $M$ be a compact oriented smooth manifold of dimension $n$
 with boundary $\partial M$ (possibly empty) and let 
$\partial M$ be given the induced
 orientation \cite{spivak0,spivak1}. 
For a $(n-1)$-form $\phi$, we have
\begin{equation}
\int_{M} d \phi = \int_{\partial M} \phi
\end{equation}
The integral $\int_{{\cal C}} \omega$ of a k-form 
 is said to be path-independent if the 
value of this integral depends
 only on the boundary $\partial{\cal C}$
 of the oriented $k$-chain ${\cal C}$. This  implies that
 $\int_{{\cal C}} \omega =0$ for every closed
$k$-chain. This property can be used to state the important result:
 a  form $\omega$ defined on a manifold $M$ is exact iff
  $\int_{{\cal C}} \omega$ is path-independent on $M$.
 For instance, 
the 1-form (\ref{angle}) is not exact. 

All these properties are generalizations of well-known results
 in vector calculus. For instance, a vector field $\vec F$ 
is conservative if 
${\vec F} = \nabla{f}$ for some function $f$ (the potential). 
Let $\omega$ be the 1-form associated to 
$\vec F$. A conservative $\vec F$ corresponds to $\omega = df$
 so that $\omega$ is exact. 
This implies that $\omega$ 
is closed and since $d\omega$ corresponds to $\nabla\times{\vec F}$,
 this implies  $\nabla\times{\vec F} =0$ as well. 
Electrostatics results from the  fact that in ${\bf R}^3$
 a  closed 1-form is exact, hence  the electric 
field is conservative. Another consequence is the widely
 used result (e.g. in thermodynamics)  that a 1-form 
$\phi =  pdx +q dy $ cannot be exact unless
 ${\partial q \over \partial x} = {\partial p \over \partial y}$.
\vskip 0.3cm

 \subsection{Topological invariants of a manifold:}

  \subsubsection{ Motivations }

 Two manifolds are said to be {\bf homeomorphic} if
 there is a continuous  mapping  from
  each other,  with continuous inverse mapping.
 A {\bf topological invariant} is a intrinsic
 characteristic of a manifold conserved by homeomorphism.
 These invariants  reveal important  features  
 and help to classify  different types of manifolds.
 Topological invariants can be numbers, scalars, polynomials,
 differential forms or more general algebraic sets such
 as groups, or algebras. Their importance for condensed matter
 Physics was recognized in a seminal paper  of Toulouse and Kleman [1]
 for the study of defects (vortices, nodal lines, textures anomalies)
 and their stability as a function of external parameters. The
 analysis of [1] depends on general characteristics of the system
 under study (e.g. dimensionality of space, nature and symmetries
 of the order parameter) and not on the precise form of the equations
 governing  the system.  Loosely speaking, different types of
 defects correspond to different (non homeomorphic) geometrical
 structures. Therefore topological invariants can help to distinguish
 between  them.
 Of course this general scheme does not tell how to compute
 the relevant invariant in a given problem. A nice example is the
 Aharonov-Bohm  effect in an infinite plane where the relevant
 invariant is the {\it winding number} (defined in the exercise 
 of section 2.2.6). An thorough  study was done in \cite{berry}.
  In the last chapter of
 these notes we discuss   two dimensional  superconductors.
 There the winding number measures the circulation of the phase
 of the order parameter around the vortices.
 In the following paragraphs we  merely give a taste of 
 the vast  subject of invariants in a manifold.

\subsubsection{ The Euler-Poincar\'e Characteristic}

 The Euler-Poincar\'e characteristic $\chi(M)$ 
  of a manifold $M$ is the oldest and the most celebrated
 topological invariant. We  shall explain how
 to calculate it for a surface. Any given   surface $S$
 can be tiled by triangles (of different sizes and shape).
 Such a partition is called a {\it triangulation}. 
 For any triangulation of $S$, denote  by $V$, $E$ and $F$
 the number of vertices, edges and faces 
 respectively. Then $\chi(S)$ is defined by
\begin{equation}
   \chi(S) =  V - E + F \,\,\, .
\label{eulpoinc}
\end{equation}
 This number does not depend on the chosen triangulation
 and is a characteristic of the surface $S$ \cite{stoker}. 
\hfill\break
  \underline{Exercise:} Show for the sphere $S^2$ that
 $\chi(S^2) = 2$. Show for the torus $T^2$ that 
 $\chi(T^2) = 0$.
\hfill\break
 For a three dimensional manifold one needs to use tetrahedra   
  to perform a `triangulation' and  the formula
 for $\chi$  becomes $ \chi =  V - E + F - T$,
 where $T$ is the number of tetrahedra. It is possible
 to generalize this notion to higher dimensional manifolds
  \cite{spivak1}.

 \subsubsection{De Rham's Cohomology}

    We have seen (section 2.2.5) that any exact form is closed
 but that the reverse is not true.
   Forms which are closed but {\it not} exact reveal  important
 topological features   and help  to  classify
  different types of  manifolds. The $k$th  {\bf cohomology
 group} of De Rham of the manifold $M$,
 $H^k(M)$,   is the set of closed $k$-forms modulo
 the exact $k$-forms.  It is a finite dimensional vector space,
 its dimension $b_k$ is called the $k$th Betti number. This integer
 is a  topological invariant  of the manifold $M$.
 A  beautiful result \cite{spivak1,greenberg}
  is that the alternating 
   sequence  of Betti numbers is the Euler-Poincar\'e
 characteristics of the manifold, namely
\begin{equation}
\chi (M) = \sum_{r=0}^{n} (-1{)^r} {b_r} 
\label{euler}
\end{equation}
 This shows that topological properties, as  $\chi$,  can
 sometimes be calculated  using analytical tools  such
 as differential forms.

 \subsection {Riemannian Manifolds and Absolute Differential Calculus}
\vskip 0.3cm

 \subsubsection{Riemannian Manifolds}

 A Riemannian Manifold is a differentiable manifold $M$ with a local
scalar product defined on each tangent space $T_{p}M$. 
 If ${\vec v}$ and ${\vec w}$ are two tangent vectors at the same
 point $p$ then
\begin{equation}
 \langle {\vec v}|{\vec w} \rangle_{p} = {\vec v}.{\vec w}
 = g_{ij}(p) v^i w^j \,\,\,\, .
\end{equation}
In particular the norm of ${\vec v}$ is given by the square-root of
       $g_{ij}(p) v^i v^j$.

 The quantities $g_{ij}(p)$  are the local components of the metric
 and are called the {\bf metric tensor}. They allow to compute
 the length ${\cal L}(\gamma)$  of a curve  $\gamma$ 
  parameterized  as in (\ref{curve}).
 We  recall that $\frac {d\gamma}{dt}$
 is the tangent vector to  $\gamma$ at the point $p = \gamma(t)$.
\begin{equation}
{\cal L}(\gamma) = \int_{-1}^{1} dt
 \left( \langle {\vec\frac {d\gamma}{dt} }|{\vec\frac {d\gamma}{dt}}
  \rangle_{p} \right)^{1/2}
=  \int_{-1}^{1}dt
\left( g_{ij}(p)\frac{d x^i}{ d t}\frac{d x^j}{ d t} \right)^{1/2}
 = \int_{0}^{{\cal L}(\gamma)} ds 
\end{equation}
 The infinitesimal arc length is denoted by $ds$. In classical books
 the metric is written as:
\begin{equation}
  ds^2 =  g_{ij} dx^i dx^j
 \label{metrique}
\end{equation}
 A {\it geodesic curve} between two points $p$ and $p'$ of $M$
 is a path of minimal length. Using variational calculus,
 one can find the Euler-Lagrange equations for a geodesic curve
 $(x_1(s),...,x_n(s))$  parameterized by its arc length $s$
 (this is a good exercise!). One obtains a system
 of differential  equations:
\begin{equation}
    \frac{d^2 x^i}{ d s^2} + \Gamma^{i}_{jk}
 \frac{d x^i}{ d s}\frac{d x^j}{ d s}  = 0 \,\,\
 \label{geod}
\end{equation}
 where the {\it Christoffel symbols} $ \Gamma^{i}_{jk}$
 are given by
\begin{equation}
\Gamma^{i}_{jk} = \frac{1}{2} g^{il}
\left( \frac{\partial g_{jl}}{\partial x^k} +
 \frac{\partial g_{kl}}{\partial x^j} -
 \frac{\partial g_{jk}}{\partial x^l} \right)
 \hbox { with }  g^{il} g_{lj} = \delta^{i}_{j} \,\,\, ,
\label{christoff}
\end{equation}
 i.e.  the matrix $g^{il}$ is the inverse of
 the metric tensor.

\subsubsection{Covariant derivative. Connexion  and curvature forms}

 We have emphasized that on a general manifold there is no way
 to compare tangent vectors at different points. However,
 given a smooth vector field,  a rather natural question to ask is
 `What is the infinitesimal variation
 of  a  vector field $\vec Y$  at point $p$ if one moves in the direction
 of  a vector $\vec V$ tangential at the point $p$ to the manifold ?' 
 A possible way out is to immerse the 
manifold in a larger space where coordinates are defined.
 But this is not intrinsic: it depends on the surrounding space.

 One needs an additional structure, called a {\bf connexion},
 that allows to compare tangent vectors  at different points
 and to differentiate vector fields (or more generally tensor
 fields). This method was invented by Levi-Civita and
 called {\it absolute differential calculus}. It is possible
 to give a general  definition of a connexion without 
 refering to the metric, as these concepts are independent.
 However we shall restrict ourselves to Riemannian manifolds.

 A Riemannian manifold can canonically be endowed with 
 a connexion provided by the Christoffel symbols
 $\Gamma^{i}_{jk}$ (\ref{christoff}).
Consider a vector field $\vec Y$ on the manifold. One defines
 its covariant derivative ${\bf \nabla}_{\vec V} {\vec Y}$ 
along the direction of the vector $\vec V$ tangential
 at the point $p$ to the manifold  by 
\begin{equation}
{\bf \nabla}_{\vec V} {\vec Y} =
\left( {\partial Y^i \over \partial x^j} V^j + 
\Gamma_{jk}^i V^j Y^k \right)e_i
\label{covarian}
\end{equation}
where $e_i = {\partial \over \partial x^i}$
 is the $i^{th}$ vector of the  basis (\ref{base})
 of the tangent  space 
$T_pM$ at the point $p$. 
 The quantity ${\bf \nabla}_{\vec V} {\vec Y}$ is a vector,
 tangent to the manifold $M$ at the point $p$. It
 represents the total  rate of  variation of ${\vec Y}$
 along the direction of ${\vec V}$. 
The first term in (\ref{covarian}) is
 nothing but the {\it convective} derivative of  ${\vec Y}$ along
 the direction of  ${\vec V}$. This term is familiar in hydrodynamics
 (e.g. in Euler and Navier-Stokes equations). 
  The second term  represents the 
 derivative of the vectors $e_i$ of the local basis of $T_pM$
  along  the direction of
 $\vec V$. Such a term is  familiar from mechanics when one uses non-cartesian
 coordinates (e.g. polar or spherical),  it reflects that
 the coordinate system  is local and point dependent.
 One can rewrite (\ref{covarian}) as follows:
\begin{eqnarray}
{\bf \nabla}_{\vec V} {\vec Y} &=&
 {\bf \nabla}_{\vec V} Y^i  e_i = 
  dY^i ({\vec V}) e_i  + Y^k {\bf \nabla}_{\vec V} e_k \,\,\, \nonumber \\
  \hbox{  with } \,\,\, {\bf \nabla}_{\vec V} e_k 
  &=&  V^j {\bf \nabla}_{ e_j} e_k  \nonumber \\  \hbox{  and } \,\,\, 
 {\bf \nabla}_{ e_j} e_k &=&   \Gamma_{jk}^i e_i
\label{localderiv}
\end{eqnarray}
 It is analogous to the well-known expression 
in mechanics of the derivative 
of a vector in the rotating frame 
$${d {\vec v} \over dt} = {d \over dt} (v^i e_i) = 
{\partial {\vec v} \over \partial t} +
 {\vec \Omega } \times {\vec v} \,\,\, .$$

\hfill\break
{\bf *Parallel Transport: }
  The formula (\ref{covarian}) for the covariant derivative 
 ${\bf \nabla}_{\vec V} {\vec Y}$  allows  to compute the variation
 of a vector field along a given direction. Inversely, from  a tangent
 vector ${\vec Y_0}$ at a point $p_0$ and a curve $\gamma$
 going from  $p_0$ to $p_1$, one can construct from 
 (\ref{covarian})  a vector field
 ${\vec Y}$ along the curve  $\gamma$ such that 
${\vec Y}(P_0) = {\vec Y_0}$. Such an operation is called {\it the parallel
 transport  of  ${\vec Y_0}$ along  $\gamma$ }. The  vector field
   ${\vec Y}$  is determined  by
$$ 0 =  {\bf \nabla}_{\vec V} {\vec Y} \hbox { with } \,\,\,
 {\vec V} = \frac{ d\gamma}{dt} \,\, . $$
 It  is therefore  obtained by solving the system
 of differential equations:
\begin{equation}
  0 =  {\partial Y^i \over \partial x^j}{ d x^j \over dt}  + 
\Gamma_{jk}^i { d x^j \over dt}  Y^k = 
{d  Y^i \over dt}  + 
\Gamma_{jk}^i { d x^j \over dt}  Y^k
\label{transp}
 \end{equation}
  with initial condition ${\vec Y}(0) = {\vec Y_0}$.
   If one considers the special case 
  ${\vec Y} = {\vec V} = \frac{ d\gamma}{ds}$ where $s$ is the
 arc-length of the curve $\gamma$, 
 equation (\ref{transp})  becomes identical to the formula  defining
 a geodesic (\ref{geod}). The equation of a geodesic
 is thus $ {\bf \nabla}_{\vec V} {\vec V}=  0 \hbox  { with } 
 {\vec V} = \frac{ d\gamma}{ds}$. In other words a geodesic
 is a  curve which is parallel transported along itself, as this  is 
  well-known  for straight   lines in Euclidean spaces. 

        \hfill\break
{\bf Remark:} We did not give any explanation
 for the appearance of the Christoffel symbols in
 the  formula  (\ref{localderiv}) for the covariant derivative
 of the vectors of the basis. This is in fact the heart of
 Levi-Civita's construction. We encourage the reader to develop
 an intuition on this formula through his/her  own readings
 \cite{spivak2,stoker,milnor2}.
  A  nice (and historical)  approach is provided by
 surface theory in ${\bf R}^3$.
  A  tangent vector ${\vec V}$ at  a point
 $p$ of the surface  is transported  to an
 infinitesimally  close point $p'$ by first  bringing 
 it parallel to itself from $p$ to $p'$. As  ${\vec V}$ 
 has no reason to be tangent  to the surface
 at $p'$, one projects ${\vec V}$
 on the tangent plane at $p'$ and obtains a new vector 
 ${\vec V'}$ tangent at $p'$. 
 The variation of ${\vec V}$ through this operation provides 
 the formula for  the covariant derivative (this is explained in
  \cite{stoker}). Another approach
 is via  `Tensor Calculus':  the first term in the r.h.s.
 of (\ref{covarian}) is not a tensor,
 so one has to add to it something
  to have  a tensorial   derivative.
 The Christoffel
 symbols are not tensors themselves but  adding
 them to the  first term turns  the sum  into  a tensor. 
 This  approach is well
 explained in \cite{kreyszig}, or in books about General Relativity 
 \cite{schutz,schrodinger}.

  \hfill\break

  We now  give a more compact expression for the covariant derivative.
 One first notices that ${\bf \nabla}_{\vec V} {\vec Y}$ 
is linear in ${\vec V}$, i.e.  ${\bf \nabla} {\vec Y}$
   can thus be viewed as a linear operator
  which  acting on   ${\vec V}$
 produces another  tangent vector at $p$. Equivalently,   
 ${\bf \nabla} {\vec Y}$ is tangent-vector valued  1-form
 and can be written  as 
$$  {\bf \nabla} {\vec Y} =
 \left( d \,\,  Y^i  + \Gamma_{jk}^i dx^j \,\,  Y^k  \right)e_i
    \,\,\, . $$
 From this expression we see that 
  ${\bf \nabla}$ itself acts as   a
 linear operator on  the  vector  ${\vec Y}$. 
 Recalling   that a matrix $L$ acts on  ${\vec Y}$ 
 as  $L{\vec Y} = L^{i}_{k} Y^k e_i$, we    write:
\begin{equation}
{\bf \nabla} = d + \omega 
\label{synthet}
\end{equation}
 where $\omega$ is a  matrix  whose coefficients are 1-forms. Precisely 
$\omega$ has matrix elements
\begin{equation} 
 \omega^i_k = \Gamma_{jk}^i  dx^j \,\,\, .
\label{connec}
\end{equation}
 The matrix-valued 1-form   $\omega$
  is called the {\bf  connexion} 1-form  and $\bf \nabla$
  the {\bf  covariant derivative.}
The {\bf curvature} $\Omega$ of a Riemannian Manifold  is defined
 by the matrix valued 2-form
\begin{equation}
\Omega = d \omega + \omega \wedge \omega
\label{curvature}
\end{equation}
i.e. for two vectors $\vec X$,$\vec Y$ on the manifold,
 $\Omega ({\vec X}, {\vec Y})$ is a matrix. Notice that 
in general the connexion 1-form  can be written as 
\begin{equation}
\omega = A_j dx^j 
\end{equation}
where  the $A_j$'s  are matrices. 
Then, $\omega \wedge \omega = \sum_{i<j} dx^i \wedge dx^j$ is non zero when
 the matrices do not commute. 
\vskip 0.3cm
\underline{Exercise:} Find in a book the formula of the Riemann
 curvature tensor (see e.g. Spivak \cite{spivak2}, Vol II,
 Chap 4, p189) in terms of
 the Christoffel symbols.   Using (\ref{connec}), verify  that 
 $\Omega$ defined in (\ref{curvature}) is  indeed the Riemann tensor.
  
\subsection{*The Laplacian}
\vskip 0.3cm 
    We saw that the spaces  of $k$-forms and $(n-k)$-forms
    over a manifold of dimension $n$ have the same  dimension.
    It is possible to define a 
     duality transformation $*$,  called {\it the Hodge star,}
  between these two spaces.
      In the Euclidean  space ${\bf R}^3$ the Hodge
 duality is given by:
   \begin{eqnarray}
    *1 &=&  dx \wedge dy \wedge dz   \nonumber  \\
      {*}dx  =  dy \wedge dz \,\,\, , \,\,\, 
  *dy &=& dz \wedge dx  \,\,\, , \,\,\,   *dz=dx \wedge dy,
          \nonumber  \\
      {*}(dx \wedge dy)  =  dz  \,\,\, , \,\,\, 
      {*}(dy \wedge dz)  &=&  dx  \,\,\, , \,\,\, 
      {*}(dz \wedge dx)  =  dy 
       \nonumber  \\
       { *}(dx \wedge dy \wedge dz) &=& 1  
      \end{eqnarray}   
       More generally the duality $*$  operation on
 a manifold {\it  depends on the local metric}, and is given by:
 \begin{eqnarray}
 *(dx^{i_1} \wedge dx^{i_2} \wedge ...\wedge  dx^{i_k}) = |g|^{1/2}
    dx^{i_{k+1}} \wedge ...\wedge  dx^{i_n}
\end{eqnarray}
where $(i_1,i_2,...,i_k,i_{k+1},...,i_n)$ is a permutation
 of $(1,2,...,n)$ with positive signature and $|g|$
 is the determinant  of  the metric $(g_{ij})$ (i.e.
 $|g|^{1/2}$  is 
 the volume element). If $\phi$ is a $k$-form, one can check
 (exercise!) that 
 $ **\phi = (-1)^{k(n-k)}\phi \,\,\, .$

    The Hodge duality  allows to define
      a scalar product in the space of k-forms by
     \begin{equation}
      (\phi_k, \psi_k) = \int_{M} \phi_k \wedge {*}\psi_k 
      \label{prodform}
     \end{equation}
     where the integral is over the manifold. \hfill\break
 \underline {Exercise:} Show that
  the scalar product of two 1-forms 
      $\phi = A_1 dx + B_1 dy +C_1 dz$ and
       $\psi = A_2 dx + B_2 dy +C_2 dz$ in ${\bf R}^3$ is 
          $$ (\phi, \psi) = \int_{R^3} (A_1A_2 + B_1B_2 + C_1C_2) dxdydz.$$
 The exterior derivative $d$ maps
 $k$-forms onto $(k+1)$-forms. One  defines  the 
adjoint exterior derivative $\delta$ which maps $k$-forms
 onto $(k-1)$-forms via
$ (\phi_k, d \psi_{k-1}) = ( \delta \phi_k,  \psi_{k-1}) \,\, .$
\hfill\break
 \underline {Exercise:} \hfill\break
  1.  Show by integrating  by part \cite{remarque} that
  in a $n$-dimensional vector space 
 $$\delta = (-1{)^{nk + n +1}} *d{*} \,\,\, . $$
 2. Deduce that   $\delta ^2 =0$.  A
  {\it  co-exact}  $k$-form $\omega$  satisfies  the global condition
 $\omega = \delta\phi$ where $\phi$ is
 a $(k+1)$-form.  A {\it  co-closed }  
  $k$-form $\omega$ satisfies $\delta\omega = 0 \,\,\, .$ 

\hfill\break

The {\bf  Laplacian} is an operation which
 takes $k$-forms onto $k$-forms
  and  generalizes  the usual Laplacian on functions.
It is defined by
\begin{equation}
\triangle = (d + \delta {)^2} = d \delta + \delta d
\label{laplac}
\end{equation}
 \hfill\break 
 \underline {Exercise:}
 In ${\bf R}^3$ show that  if $f$ is a function, one has  
 $\triangle{f} = -\left( {\partial_x}^2 f +
 {\partial_y}^2 f + {\partial_z}^2 f \right)$.

A $k$-form $\omega_k$ such that
 $\triangle \omega_k =0$ is called {\bf  harmonic.}
 For a smooth enough $k$-form, a necessary 
and sufficient condition for harmonicity is to be closed and co-closed:
$$\triangle \omega_k =0 \Longleftrightarrow d \omega_k =0 
 \hbox { and } \,\,\,    \delta \omega_k =0 \,\,\, .$$
{\bf   Hodge's  theorem:} On a compact manifold without boundary
 any $k$-form $\omega_k$ can always 
be decomposed into the sum of an exact form $ d \alpha_{k-1},$
 a co-exact  form $\delta \beta_{k+1}$   and a harmonic form
 $\gamma_k$ \cite{warner,rosenberg}:
\begin{equation}
\omega_k = d \alpha_{k-1} + \delta \beta_{k+1} + \gamma_k
\label{hodge}
\end{equation}
 This very important result is well-known in $3d$
 vector analysis 
of electromagnetism of continuous media as the 
Helmholtz decomposition according to which for a closed and 
compact manifold or for a controlled growth
 of the fields at infinity, any smooth vector field $\vec B$ can 
be decomposed as $\vec B = \nabla \phi + \nabla\times \vec M + \vec H$,
 where $\vec H$ has both vanishing curl and divergence.

\hfill\break   Remark: 
 if $\omega_k$ is closed 
 then $d \delta \beta_{k+1} =0$, which implies, as  $d$ and $\delta$
 are adjoint,
 that $\delta \beta_{k+1} =0$.
 Hence  $\omega_k = d \alpha_{k-1} + \gamma_k \,\, .$ 
 Therefore $\omega_k$ and $ \gamma_k$ belong to the same
 cohomology class. Denoting  by  Harm$^k (M)$  the 
 space of harmonic $k$-forms on $M$,
 the Hodge theorem establishes an isomorphism between the two spaces 
$H^k (M)$ and Harm$^k (M)$ and according to (\ref{euler}),
  the Euler-Poincar\'e characteristic is given by 
\begin{equation}
\chi (M) = \sum_{k =0}^{n} (-1{)^k}\hbox{dim Harm}^k (M)
\end{equation}
 This relation shows that 
 properties  of the Laplacian  probe the   topology 
  of  the manifold $M$ \cite{rosenberg}.

  \subsection{Bibliography}

  We do not intend to be exhaustive, we just list 
    books  we used  and give some
 (subjective) comments.
  A classical and outstanding introduction to  Geometry
  in general is \cite{hilbert}. A `modern-classic'  more
 inclined toward topology and analysis is \cite{milnor1}.
  A good way to  start  with differential geometry is to study 
  curves and surfaces. 
  A  very pleasant book  on these  topics, full of results, figures
 and historical comments is  \cite{struik}. 
 Some books   begin with curves and surfaces and then
 introduce the general concept of $n$ dimensional manifolds.
 This can be  a very helpful   point of view 
 since  it is not too  formal 
  \cite{stoker,kreyszig,mccleary}. More  recent  works  
 are  \cite{dubrovin} with emphasis on
  applications, \cite{warner} more inclined
 towards algebra  and \cite{rosenberg} devoted to analysis
 on manifolds and  well adapted to theoretical
 physicists. An exhaustive treatment is to be found in
 \cite{spivak0,spivak1,spivak2} and the following volumes.
 We  owe  a lot to the review paper \cite{eguchi} 
  written for   high energy physics  and Yang-Mills
 theories. For a pedagogic  introduction to differential forms,
 see \cite{flanders}.

 \section {Fiber bundles and their topology}
\vskip 0.3cm

 \subsection{Introduction}

 The concepts of connexion and curvature that we have defined for 
 Riemannian manifolds can be extended  to a more general structure
 called a  {\bf   fiber bundle}.
  A  fiber bundle is  a manifold ${\bf X}$ 
   that locally looks like the product of two simpler manifolds.
 For instance the torus ${T^2} = {S^1} \times {S^1}$ is globally
 the product of two  1-dimensional circles ${S^1}$.
 A cylinder is also a global  product  ${S^1} \times [-1,1]$.
 A   M\"{o}bius strip is locally the  
 product  of an  arc  of  ${S^1}$ by $[-1,1]$ but not globally.
 More precisely, a fiber  bundle is a triplet $({\bf X}, M, \pi)$
 where ${\bf X}$ is a manifold ({\it the total space}), $M$
 ({\it the base space})  a submanifold
  of ${\bf X}$  and $\pi$ ({\it the projection}) a smooth function
 from  ${\bf X}$  to $M$:
 \begin{eqnarray}
 \pi:    {\bf X}  &\to& M \nonumber \\
         x  &\mapsto&  \pi(x) = p 
\end{eqnarray}
 The inverse image $\pi^{-1}(p)$ of any point $p \in M$
 is called the {\it fiber} above $p$ and is isomorphic
 to a given manifold  $F$.
 Intuitively, a fiber bundle is a collection of identical 
(isomorphic) manifolds $F$  which depend  on a parameter $p$
 belonging to the base  manifold $M$. When  the fiber  $F$
 is a vector-space ${\bf R}^n$ or  ${\bf C}^n$
 the bundle is called a {\it vector-bundle}.
(see e.g.  \cite{husemoller}).

   A  {\bf section} $s$  of a fiber  bundle is a smooth function
  which associates to each point $p  \in M$
 a element of the fiber above $p$. Therefore one has
  $\pi\circ{s} = Id_M$
 where $Id_M$ is the Identity function in $M$.

\hfill\break
{\it Examples: }\hfill\break
1. The product $M_1\times{M_2}$
 of two manifolds $M_1$ and $M_2$ is a fiber bundle called
 the trivial bundle.
 One can take either $M_1$ or $M_2$ to be the base
 and the other one to be the fiber. \hfill\break
 2.  The total  tangent space of $M$ is a vector  bundle also  called
  the {\it tangent bundle}.
 The base space  is the manifold  $M$ itself and the fiber
 above the point $p$ is $T_pM$ which is homeomorphic to
${\bf R}^n$. The projection is exactly the one defined
 in section 2.2.1. And  what is a section of the tangent 
 bundle? A vector-field!\hfill\break
 3. More generally the space of $k$-forms 
   is also a vector bundle over the manifold  $M.$
 \vskip 0.2 cm

 The fiber  bundles  described here are  assumed to be
 {\it locally trivial: }   for any
 point $p$ of the base $M$ there is a neighborhood
 $U_p$ such that the restriction of the  bundle over  $U_p,$ 
  i.e. $\pi^{-1}(U_p),$  is homeomorphic to $U_p\times{F}$.

  \subsection{Local symmetries. Connexion and Curvature}

   In physical applications one considers vector bundles
  where  the base space  
   is the physical space and   the fiber 
 is a  representation  of a continuous symmetry group, i.e.
 a Lie group ${\bf G}$ with Lie algebra ${\cal G}.$
  At each point, the order parameter is an element of the fiber
  over that point  and a local  symmetry group 
 acts upon it. The order parameter is precisely  a section
 of a  vector  bundle. It is natural   to ask
 about   the variation  of the order parameter when one
 moves from one point of the base to another.
  However, although  all fibers are homeomorphic  to the same 
 vector space, there is no intrinsic way to identify
 two fibers over two different  points $p$ and $p'$ of the base.
 In other words, given a section $s,$ the two vectors $s(p)$
 and $s(p')$ can not be compared a priori. This is exactly
 the same problem that we encountered  for vectors
 fields, which are sections of the tangent bundle.
 The extra-structure needed to compute the variation of a section
 $s$ along a given direction tangent to the base is called
 a  {\it fiber bundle  connexion} and is again denoted by
 ${\bf \nabla}.$  The expression of ${\bf \nabla}$
 is obtained by   reinterpreting  the formula (\ref{synthet}):
\begin{equation}
{\bf \nabla} = d + \omega \,\,\, .
\label{synt2}
\end{equation}
 Here $d$ is the usual  operation of differentiation 
 of functions and $\omega$  is a  matrix-valued  1-form
 (\ref{connec})
 on the base space $M$  that can be written 
 $$ \omega = A_i dx^i $$
 where $(x^1,...,x^n)$ is a local system of coordinates
 on the base $M$. We impose that $\omega$ represents at each point
 an infinitesimal  transformation of the group
 ${\bf G}$ i.e. the matrices $A_i$ belong to the Lie algebra
  ${\cal G}$. The {\it curvature} $\Omega$ of
 the connexion  is also given by the same formula
 as above (\ref{curvature}):
 $$ \Omega = d \omega + \omega \wedge \omega  
           =  d \omega + \frac{1}{2}[\omega,\omega]   \,\,\,  . $$
 \hfill\break
  {\bf Important examples:} \hfill\break
 1. For  the tangent bundle, we  take the
 symmetry  Lie group to be the 
  group of invertible matrices $GL(n,{\bf R})$.
 The associated Lie algebra is just the group of all
 $n\times{n}$ matrices. 
 Connexions
 on a Riemannian manifold are  a special case of 
 fiber bundle connexions. \hfill\break
  2. We  study  now
   a 2-dimensional physical
 system  $M = {\bf R}^2$ where at each point
 an order parameter (or  wavefunction) is defined and is 
a complex number $\psi = |\psi|e^{i \chi}.$ Hence  the fiber
 is  $F = {\bf C}$. The Lie group for 
 Maxwell electromagnetism is $U(1)$ and the Lie algebra is 
$i {\bf R}$. Then, the  connexion  1-form 
(or  gauge potential) can be written 
 $\omega = - i A$ where $A = {A_x} dx + {A_y} dy$, $A_x$ 
and $A_y$ being real functions ($1\times{1}$ matrices).
 Since the Lie group is abelian, the curvature  
reduces to $\Omega = d \omega = - i B$
 where  $B = (\partial_x A_y - \partial_y A_x) dx \wedge dy$
 is nothing but the magnetic field.
 The covariant derivative given by (\ref{synt2})
 is then $\nabla = d -i A$. This is indeed 
 the differential operator that appears when one studies
 the Schr\"odinger equation in a magnetic field, or
 the Ginzburg-Landau model
 for  a superconductor.\hfill\break
 3. In  the Gross-Pitaevskii description 
  for rotating  superfluid $^4 He$ 
   the connexion is the Coriolis term ${\vec \Omega} \times {\vec r},$ 
  $\vec \Omega$ being  here  the angular velocity. \hfill\break
 4. The Yang-Mills connexion is obtained by  the same construction as before
 with a non-abelian Lie group (typically $SU(2)$). The reader
 is referred  to the original paper of Yang and Mills \cite{yangmills}. 
 The  theory of fiber bundles is   the right
 setting for gauge symmetries in  high energy physics
 \cite{wuyang,viallet}.
\vskip 0.2cm

  \subsection{Chern  classes.}

 The cylinder and the Moebius strip  have both  $S^1$
 as base space and [-1,1] as fiber, but they are not
 homeomorphic. One would like  
  to classify  different types of fiber bundles
 with  given base and  fiber.
{\bf  Characteristic classes}  are  
 cohomology  classes of the base space  that are
  topological invariants
 of the  fiber bundle  \cite{milnor3}.

 We shall study the {\bf  Chern classes} for fiber  bundles
 with Lie group ${\bf G} = GL(n, {\bf C})$, the group
 of invertible complex $n \times n$ matrices.
  The Chern-Weil theorem gives an explicit 
  construction of these classes  starting 
 from a  connexion 1-form  $\omega$  and the associated
 curvature $\Omega$. This theorem also proves that
 the invariants obtained {\it  do not depend }  on the chosen
 connexion (see \cite{chern} for a most  enlightning exposition).
 
    We  recall that the curvature  $\Omega$
 is a 2-form whose coefficients are 
  $n \times n$  matrices with complex
 coefficients. A polynomial $P(\Omega)$ is 
 said {\it invariant }  if for any matrix
 $g$ in $GL(n, {\bf C})$,
 $$P(\Omega) = P(g^{-1} \Omega g) \,\,\, . $$
 For example,
det$(1 + {i \over {2 \pi}} \Omega)$ and tr$( e^{ {i \Omega} \over 
{2 \pi}})$ are  invariant polynomials.

\hfill\break {\bf Theorem (Chern-Weil):}
{\it  If $P$ is an invariant polynomial and $\Omega$  curvature  2-form, then 

(i) $P(\Omega)$ is a closed differential form ($dP =0$).

(ii) If $\omega '$ is another connexion on
 the same fiber bundle and $\Omega '$ the associated curvature, 
  then there  exists 
  a form $Q$ such that $P(\Omega ') - P(\Omega) = dQ \,\,\, .$ }
\hfill\break

   Property (i) shows that $P(\Omega)$ defines a cohomology
 class of the base space $M$ and (ii) proves that
 this  class does not depend on the chosen connexion $\Omega$.
  $P(\Omega)$ is called a {\it  characteristic class.}
 In particular, integrals  
  of $P(\Omega)$ over cycles  (i.e. submanifolds
 without boundaries)
 will  provide  topological
  invariants \cite{chern,chern2}.

 This construction can be generalized to other  Lie groups
 and the  polynomial $P$  has a  different expression according to the  
 associated the Lie 
 algebra. Here, we shall consider only the {\it  Chern  classes }
 defined from 
\begin{equation}
P(\Omega) =  \hbox{det}(1 + {i \over {2 \pi}} \Omega)
 = 1 + c_1 (\Omega) + c_2 (\Omega) +...
\label{chernpoly}
\end{equation}
where $c_i (\Omega)$ is a scalar-valued $2i$-form
 called the $i^{th}$ Chern class. 
 From the expansion  of the determinant we obtain the expressions
\begin{eqnarray}
c_0 & = &1
\nonumber  \\
c_1 & = &{i \over {2 \pi}} \hbox{tr} (\Omega)
\nonumber  \\
c_2 & = & {1 \over {8 {\pi^2}}} (\hbox{tr} \Omega \wedge \Omega 
- \hbox{tr} \Omega \wedge \hbox{tr} \Omega )
\label{chernex}
\end{eqnarray}   
and $c_i =0$  when  $2i$ is greater than the dimension
 of  $M$.
Since $dP =0$,  each 
Chern  form is closed as well: $d c_i (\Omega)=0 \,\,.$
 The $i^{th}$ Chern  class   belongs to the 
cohomology group $ H^{2i} M \,\,.$
 A  remarkable fact is
 that the Chern forms define 
{\it integer}  cohomology classes:   the integral of
 $c_i (\Omega)$ over any $2i$-cycle ${\cal C}$
 (i.e.  any  oriented submanifold ${\cal C}$
 of $M$, of dimension $2i$ and without boundary)
 is an integer, depending upon ${\cal C}$ but   independent 
 from  the  connexion  1-form  $\omega$ over the fiber bundle:
\begin{equation}
\int_{{\cal C}} c_i (\Omega) = n \in {\bf Z}
\label{chernc}
\end{equation}
 This integer $n$ is a topological invariant of $M$
 called a {\it  Chern number.}
 
\hfill\break
Remark:  if $P(\Omega)$ is  homogeneous  of degree $r$
 (see e.g. (\ref{chernex})) an explicit
 formula  can be given  for the form
 $Q$ that appears in the Chern-Weil
 theorem:
\begin{equation} 
Q = r  \int_{0}^{1} dt P( \omega ' - \omega, {\Omega_t},...,{\Omega_t})
\label{clasb}
\end{equation}
where $\Omega_t$ is the curvature associated to  the
 interpolating connexion ${\omega_t} = t \omega ' + (1-t) \omega$.

\hfill\break 
{\it Examples:}\hfill\break 
1.  Consider the $U(1)$ abelian 
bundle described by the curvature $\Omega = -i B$ where $B$ is 
the  magnetic field. 
 On  a two dimensional  base space,
 we have from (\ref{chernpoly})
$$P(\Omega)  
  = 1 + {B \over {2 \pi}}$$
 and the only non-zero Chern class is
 $c_1 (B) = {B \over {2 \pi}}$. Then, if $M$ is a closed
 surface without boundary
\begin{equation}
{1 \over {2 \pi}} \int_{M} B = n \in {\bf Z}
\label{flux}
\end{equation}
corresponds to a  well-known flux quantization condition \cite{sakurai}.
 \hfill\break 
2.  More generally, on a manifold $M$, all the $k$-fold 
exterior products $\Omega, \Omega \wedge \Omega,\Omega \wedge
 \Omega \wedge \Omega$ etc... 
$(k \leq {1 \over 2}\hbox{dim}M)$ lead to topological invariants.
 If dim$M =4$, only the two first are 
relevant
\begin{eqnarray}
\int_{M} \Omega \propto \oint \vec B .
 \overrightarrow{dS} & = & n \in {\bf Z}
\nonumber  \\
\int_{M} \Omega \wedge \Omega \propto \int \vec E . \vec B d^4 x 
& = & m \in {\bf Z}
\end{eqnarray}   
\hfill\break

To end this section, we discuss the issue of
 what makes certain physical quantities topological numbers
 and some others not. To that 
purpose, we consider two examples taken from geometry and electromagnetism.
\hfill\break
{\it Example 1:} 
Consider the 2-sphere $S^2$ of radius $R$. Its curvature
 is $ K = \frac{1}{R^2}$.
 Then the area of the 
sphere $\int_{S^2} dS = 4 \pi R^2$ is a  metric  invariant.
 But $ \int_{S^2} \Omega = \int_{S^2} K dS = 4 \pi$
 is a topological invariant and is  conserved by  smooth 
deformations of the sphere that change the metric.
\hfill\break 
{\it Example 2.} 
For electromagnetism, the electric charge inside
 a closed surface $S$ is given by 
$${ 1 \over 4 \pi} 
\int_{S} \vec E. \overrightarrow{dS} = 
{ 1 \over 4 \pi} \int \nabla .  \vec E dV \,\,\, .$$
 This statement is equivalent  to 
$\nabla.\vec E = \rho$ (charge density).
 The conservation of the electric charge 
in time is $ {d \over dt} \oint_{S} \vec E.\overrightarrow{dS}
=0$. It is equivalent  to the Maxwell-Faraday equation.
 These two relations do not 
refer to topological invariance although it is
 sometimes asserted that a topological invariant is obtained by 
integrating a total divergence. The electric charge 
is a counterexample of this statement.

\vskip 0.3cm 
 \subsection{Manifolds with a boundary: Chern-Simons classes.}
\vskip 0.3cm 

 \subsubsection{The Gauss-Bonnet theorem }

 We recall some facts concerning a
  surface  $S$  embedded in ${\bf R}^3$ \cite{struik}.
 To investigate geometrical  properties, Euler had 
 the idea to study curves drawn on $S$.  More precisely,
 he considered {\it normal sections} at a point $p$,
  i.e. plane  curves obtained
 by intersecting the surface $S$ with its normal plane
 at a point $p$. Euler proved that such 
 all normal sections have a curvature $\kappa$ at $p$
  limited by two extremal values $\kappa_1$
 and  $\kappa_2$ called {\it principal curvatures}.
 Moreover, if  the  principal curvatures are not
 equal, there is one curve $C_1$ having curvature $\kappa_1$
 and another $C_2$,  perpendicular to the first one, having
 curvature $\kappa_2$. If a normal section makes 
 an angle $\phi$ with  $C_1$, its curvature is
 given by $\kappa = \kappa_1 \cos^2\phi  + 
 \kappa_2\sin^2\phi  \,\,.$

\hfill\break \underline{Exercise:} Prove this theorem
 of Euler (hint: write the local equation of the surface as 
 $z = z(x,y)$ and Taylor-expand to the second order
 \cite{struik,spivak2}).

\hfill\break

 The product of the two principal curvatures $\kappa_1$
 and  $\kappa_2$ at a point $p$ is denoted
 by $K = \kappa_1\kappa_2$. The scalar  $K$ is the {\bf Gaussian
 Curvature} of the surface $S$ at the  point $p$. The Gaussian
 curvature is conserved  under   deformations that preserve
 the metric (e.g. bending a surface). It is a metric invariant,
 that only depends on the intrinsic geometry of the surface
 and not on the surrounding  space, although the above definition
 involves this embedding space. Gauss proved  this fact
 by discovering an explicit formula for  $K$ in terms 
 of the local metric $g_{ij}$  and its derivatives.
 This is the {\it Theorema Egregium} (1827), a most `remarkable
 theorem' \cite{struik,stoker,kreyszig}.
 
\hfill\break {\it  Suggestion:} Do not try
 to prove the  Theorema Egregium.
\hfill\break \underline{Exercise:} Check that the Gaussian
 curvature of a plane is 0.
 Prove that the Gaussian
 curvature of the sphere $S^2$ of radius $R$ is $ K = 1/R^2\,\,.$
 Since $S^2$ and the plane have different Gaussian curvatures
 it is not possible to make an  isometric mapping, even local,
 between them. There is no faithful  plane  map of the Earth
 that would respect the distances.

\hfill\break

  The {\bf Gauss-Bonnet theorem} \cite{stoker}  for a closed
 surface $S$  establishes a relation between  the curvature,
 which is a metric invariant and  the Euler-Poincar\'e
 characteristic of  $S$:
 \begin{equation}
  \frac{1}{2\pi} \int\int_{S} K dS  = \chi(S)  \,\,\, .
 \label{gaussbonnet}
 \end{equation}
 
 The quantity $ \frac{1}{2\pi} \int\int_{S} K dS$
 where $dS$ is the surface element, is sometimes  called 
{\it  curvatura integra}. The Gauss-Bonnet theorem 
 establishes that the curvatura integra is
 a topological invariant of closed surfaces.
 It has numerous and profound implications.
 For example  \cite{kreyszig},
 as  the reader can check, 
 it implies that no metric of negative curvature
 can be defined on sphere, or that no metric
 with strictly positive or strictly  negative curvature
 can be defined on a torus (for a torus, $\chi= 0).$

 One should notice the similarity between
  the theorem of Gauss-Bonnet
(\ref{gaussbonnet}) and the Chern numbers (\ref{chernc})
 defined as topological invariants obtained by integrating
 a Chern class. The curvature $K$ of a surface can   be
 defined as a characteristic class (called the  Euler class).
 So in fact these two expressions are not only similar:
 they have the same origin. Chern developed the theory
 of characteristic classes while studying   higher dimensional
 generalizations of the Gauss-Bonnet theorem \cite{chern45}.

  \subsubsection{Surfaces with boundary}

  We now consider  that the surface $S$ has a boundary and is
 {\it oriented}. An orientation means  that
 an outward  unitary  normal vector ${\vec N}$  can be defined
 in  a coherent manner throughout the surface
 (for instance this is not possible for a Moebius strip).
 
  Let $\gamma(s)$ be a  space curve  parameterized by its arc-length
 $s$, with  $ds^2=dx^2+ dy^2+dz^2.$  Suppose that 
 $\gamma$ is drawn on  the surface $S$.
  To  each point $p\in{S}$ on  the curve,
 one associates a local, orthonormal and direct frame
 $({\vec t},{\vec u},{\vec N})$ where ${\vec t} = \frac{d\gamma}{ds}$
 is a unitary tangent vector to the curve $\gamma$, ${\vec u}$
 is perpendicular to ${\vec t}$ and belongs to the tangent plane
  of $S$ at $p$ and   ${\vec N}$ is the normal vector at $p$.
 One has in particular
  $$ {\vec u} = {\vec N} \times {\vec t} \,\,\, .$$
 The curvature vector ${\vec k}$ of the curve $\gamma$
 is defined by
 \begin{equation}
 {\vec k} =  \frac{d {\vec t} }{ds} \,\,\, .
 \end{equation}
 As ${\vec t}$ is unitary, the curvature vector ${\vec k}$
 has components on ${\vec u}$ and ${\vec N}$. 
 Its projection on  ${\vec N}$ is called the {\it normal
 curvature} of $\gamma$, whereas the projection on
  ${\vec u}$ defines the {\it geodesic curvature}  $k_g$:
 \begin{equation}
 k_g = {\vec k}. {\vec u} =  \left( {\vec t}\times \frac{d {\vec t} }{ds}
 \right).{\vec N}
\label{geodes}
 \end{equation}
The geodesic curvature
 $k_g$ vanishes for a geodesic  curve of the surface $S$
 \cite{struik,kreyszig}. 
 We can now give the generalization
 of the Gauss-Bonnet theorem if the 
  the boundary $\partial S$ is not empty. As  the
 curvatura integra is not  an integer in general,
   a boundary contribution must be added \cite{stoker}:
\begin{equation}
  {1 \over {2 \pi}}
\int {\int_S} K dS + {1 \over {2 \pi}}  \oint_{\partial S} {k_g} dl
 = \chi (S) 
\label{gb}
\end{equation}
where $K$ and $k_g$ are respectively
 the curvature  of the manifold $S$
 and the geodesic curvature  of its  boundary.

\hfill\break 
{\it Example:}  A spherical cap  in the north hemisphere,
 containing  the north pole  and limited by 
 circle of latitude  $\theta_0$, 
is topologically equivalent  to a disk i.e. to
a triangle. Hence  its Euler-Poincar\'e characteristic   
 is equal to  1. 
 The geodesic curvature of the 
 circle of latitude  $\theta_0$
  is ${k_g} = { 1 \over R} \hbox{cotan}  {\theta_0}$. 
 The  Gauss-Bonnet theorem (\ref{gb}) gives indeed 
\begin{equation}
{  \hbox{Area of the cap}  \over 2 \pi R^2} 
+ {1 \over 2 \pi} 2 \pi R {k_g} \sin {\theta_0}  =1
\end{equation}
For ${\theta_0} = {\pi \over 2}$, i.e. at the equator (half-sphere case),
 the boundary is a geodesic and $k_g =0$ so that only the 
first term in  (\ref{gb})  contributes.

\subsubsection{Secondary characteristic classes}

 We showed that the Gauss-Bonnet theorem need to
 be generalized  to incorporate the contribution of a boundary. 
 Is it possible to  modify the characteristic classes theory
 accordingly?
The answer is again  contained in  the
 Chern-Weil theorem. 
 Since $P(\Omega ') - P(\Omega) = d Q$,
 the integral of $d Q$ vanishes if the manifold has no boundary.
 But if there
 is a boundary $\partial M$, then by Stokes' theorem
\begin{equation}
\int_{M} dQ = \int_{\partial M} Q
\end{equation}
 needs not be zero 
and  corresponds  precisely the 
contribution of a  `geodesic curvature'.
 Hence  (\ref{chernc})  rewrites \cite{chern2}:
\begin{equation}
\int_{M} c_i (\Omega) - \int_{\partial M}
 Q(\omega, {\omega_0}) = n \in {\bf Z}
 \label{simon}
\end{equation}
where $\omega$ is the  connexion  1-form associated 
to the curvature $\Omega$ and $\omega_0$
 is a well chosen  connexion  that compensates
 boundary effects.  The forms $Q$ 
 associated to each Chern forms $c_i$ are
 called {\bf Chern-Simons  classes } 
 or {\it  secondary}  characteristic  classes.
 We  now present some explicit calculations.   The 
 characteristic  class 
associated to $\Omega = d \omega + \omega \wedge \omega$ 
 for  the Yang-Mills case  is
 $\hbox{tr}(\Omega \wedge \Omega)$. Consider 
another connexion $\omega '$ and the interpolation
 between $\omega$ and $\omega '$ defined by 
\begin{eqnarray}
\omega_t & = & t \omega ' + (1-t) \omega \\
\nonumber
\Omega_t & = & d \omega_t + \omega_t \wedge \omega_t
\end{eqnarray}
for $t \in [0,1]$. Define $\alpha = \omega - \omega '$
 so that by (\ref{clasb}), $Q$ is given by
\begin{equation}
Q(\omega, \omega ') = 2 \int_{0}^{1} 
 \hbox{tr}( \alpha \wedge \Omega_t) dt
\end{equation}
Since $\Omega_t = \Omega - t d \alpha + t^2 \alpha \wedge \alpha
 - t \alpha \wedge \omega - t \omega \wedge 
\alpha$, then 
$$Q (\omega, \omega ') = 2 \hbox{tr} ( \alpha \wedge \Omega -
 { 1 \over 2} \alpha \wedge d \alpha 
+ {1 \over 3} \alpha \wedge \alpha \wedge \alpha -
 {1 \over 2} \alpha \wedge \alpha \wedge \omega 
-{ 1 \over 2} \alpha \wedge \omega \wedge \alpha) \,\,\,.$$
 The cyclicity of the trace together with the 
 property (\ref{sign})  of the wedge-product  provides 
$$ Q (\omega, \omega ') = 
 \hbox{tr} (2 \alpha \wedge \Omega - \alpha \wedge d \alpha 
- 2 \alpha \wedge \omega \wedge \alpha 
+ {2 \over 3} \alpha \wedge \alpha \wedge \alpha ) \,\,\,.$$
Taking $\omega ' =0$ we recover
 a famous  expression for the Chern-Simons connexion 
\cite{jackiw} namely    
\begin{equation}
Q (\omega) = \hbox{tr}( \omega \wedge d \omega + {2 \over 3}
  \omega \wedge \omega \wedge \omega ) 
\end{equation}
When applied to the abelian connexion U(1)
 on a domain $M$  of the 2d plane, we obtain
$Q( A, A') = A -A'$ and the curvature
 is related to the magnetic field $\Omega = - i B$.
 The vector potential 
$A'$ to be equal to $ \nabla \chi$, where
  $\chi$ is the phase of the order parameter
 $\psi = |\psi| e^{i \chi}$. 
 Then, according to (\ref{simon}) we have
\begin{equation}
\int_{M} B + \oint_{\partial M} ( \nabla \chi -A ) =n \in {\bf Z}
 \label{fluxo1}
\end{equation}
which corresponds to the well-known fluxoid expression.
 By analogy with the Gauss-Bonnet theorem for a surface
 with boundary (\ref{gb}), we can say that $B$ is a curvature
 and that $( \nabla \chi -A )$, which we shall
 interpret in section 5,  plays the role
 of a geodesic curvature.

\vskip 0.3cm 

 \subsection{*The  Weitzenb\"ock formula}

The  definition of the Laplacian on forms (\ref{laplac})
is  not always  adapted  to describe problems we 
aim to solve. For a quantum 
 particle moving in a magnetic field described by 
 a U(1)-connexion or for  the related Ginzburg-Landau equation,
 the Hamiltonian $H$  (or the free energy) is given 
 in terms of the covariant derivative $\nabla = d - iA$ 
and of its adjoint $\nabla^*$ with respect to 
the scalar product on 1-forms  defined in (\ref{prodform}), e.g. 
  $ H = {1\over 2} \nabla^* \nabla $ 
in units where both the mass of the particle and $\hbar$
 are set to one. Thus  the question arises to relate 
the covariant derivative to the Laplacian.

 We first denote by $D$ the operator 
$D = d + \delta$ so that the Laplacian
 $\Delta = d\delta + \delta  d = D^2 $.
 On a flat space and in the dual 
basis $\epsilon_i = dx^i$, we have 
$D f = \sum_{i} \epsilon_i \partial_i f$ so that
 $D^2 f = - \sum_{i} \partial_i^2 f$ coincides 
with the usual Laplacian on functions.
 But more generally for a Riemannian manifold
 a curvature term, the local basis 
vectors $\epsilon_i$ are also dynamical variables
 so  we   write 
  $Df = \sum_{i}\epsilon_i   \nabla_i f$, 
where $\nabla_i$ is the
 covariant derivative along the $i$-direction, then, 
$$ D^2 f = - \sum_{i} \nabla_i^2 f +
 \sum_{j < i} \epsilon_j   \epsilon_i
  (\nabla_j \nabla_i - \nabla_i \nabla_j)f$$
The second term in the rhs corresponds
 precisely to the curvature $K$ so that
 we have the general expression
\begin{equation}
D^2 f = \nabla^* \nabla f + K f
\label{W}
\end{equation}
known as the Weitzenb\"ock formula \cite{rosenberg,roe}.
 
  Coming back to the example of a quantum 
 particle in a magnetic field, 
 we can identify the curvature $K$ with the magnetic field.
 For the case of a constant magnetic field, the $K$-term 
 in (\ref{W}) is just a constant so that the Hamiltonian
 coincides with the Laplacian. Therefore, on a compact 
 manifold without boundary,
 the  Schr\"odinger equation  is geometrical and admits 
 topological invariants, namely the total quantized magnetic flux.
 This  is not true anymore in the presence of boundaries. 
 For  a  non-uniform magnetic 
 field,  the curvature $K$ becomes
 a local function of the coordinates 
  and  the problem defined by the Hamiltonian 
 $ H = {1\over 2} \nabla^* \nabla $   has no   
  geometrical features. To recover a geometrical formulation,
 Aharonov and Casher 
 \cite{ac} proved  that  a  Zeeman term
  (${1 \over 2} \vec \sigma.\vec B$), 
  playing  the  role of $K$ in  (\ref{W}), 
 must be added. Similar non-geometrical features appear
  for  the Pauli or Dirac equations. These equations 
  admit  zero-modes solutions
  of the type  $D f =0$ 
  whose   number $n$ is  an    Atiyah-Singer 
  Index. But  this number  {\it  does not}  correspond to 
   a Chern number of the type (\ref{flux}) since 
  the magnetic field is not a dynamical variable 
  of the problem but only a parameter and   its magnetic flux
 is not  quantized.     
  To obtain an example of a U(1)-connexion  
  with topological invariants,
  we consider now the case of a two-dimensional superconductor
 described by the Ginzburg-Landau equations, where both
 the order parameter and the magnetic field are dynamical variables.

\vskip 0.3cm

 \section{The dual point of Ginzburg-Landau equations
 for an infinite system}
\vskip 0.2cm

 The existence and  
stability of vortices in superfluid or superconducting
 systems have been mainly studied for 
the case of infinite systems or in a limit 
where boundary effects do not play an essential role.
 Among the large variety of 
methods available  to study vortices in superfluids
 or superconductors, we choose to work 
in the framework of the Ginzburg-Landau expression for the free energy. 
We consider the case of a finite $2d$
 superconducting bounded domain (a billiard) and 
study the existence and stability of vortices. The superconducting state 
is characterized by a complex order parameter.
For infinite systems, nonlinear functionals given by  
Ginzburg-Landau, Gross-Pitaevskii or Higgs expressions
 admit vortex like solutions.
 These solutions are characterized by topological numbers e.g. the 
number $n$ of vortices. How can  these results 
  be extended to finite size systems? 
 Is there a mechanism by which boundary conditions
 may allow to select a state with a given 
 number of vortices ?

\vskip 0.3cm

 \subsection{The Ginzburg-Landau equations}
\vskip 0.3cm

  The Ginzburg-Landau equations
  describe a superconducting billiard if  both the order 
     parameter and the vector potential have a slow spatial variation.
  The expression of the Ginzburg-Landau
 energy density $a$ is 
\begin{equation}
a= {a_0} + {a_2}|{\psi}{|^2} + {a_4}|{\psi}{|^4} +{a_1} 
 |({\vec{\nabla}}- i{{2e} \over {\hbar c}}
{\vec A}) \psi{|^2} + {{B^2} \over {8\pi}}
\end{equation}
where $\psi = |\psi| {e^{i \chi}}$ is the complex-valued order parameter,
 $B$ is  the magnetic field  
 and the $a_i$'s are real 
parameters. Defining \cite{degennes}
 $\xi^2 =  {{a_1} \over {|a_2|}}$,
  ${\lambda^2} = 
{1 \over {4\pi}} \sqrt 2({ {\hbar c} \over {2e} })^2 
{{a_4} \over {a_1|a_2| } }$,   
  the  dimensionless  free energy $\cal F$ is 
\begin{equation}
{\cal F} = {\int_{\Omega}} {1 \over 2} |B{|^2} + {\kappa^2}
 |1 - |\psi{|^2}{|^2} + 
|({\vec{\nabla}} - i{\vec A})\psi {|^2}
\label{gladim}
\end{equation}
  where $\psi$  is measured in units of ${\psi_0} =
 \sqrt {{|{a_2}|} \over {2{a_4}}}$, 
 $B$ in units of  
${ {\phi_0} \over {4 \pi \lambda^2}}$, and 
 the lengths in units of 
$\lambda { \sqrt 2} $. The numerical factor $\sqrt 2$
 is for further convenience.
The ratio  $\kappa = {\lambda \over \xi}$ is the only free parameter
 in (\ref{gladim}) and it determines,
 in the limit of an  
infinite system, whether the sample is a type-I
 or type-II superconductor
 \cite{degennes}. The integral is 
over the two-dimensional domain $\Omega$ of the
  superconducting sample.
The Ginzburg-Landau equations for the order parameter $\psi$ and for the
 magnetic 
field ${\vec B}= {\vec \nabla} \times {\vec A} $  are obtained
 from a variation of $\cal F$.
 They are 
  nonlinear second order differential equations.
 Their  solutions are not known 
 except 
 for some particular cases.

\vskip 0.3cm
 \subsection{The Bogomol'nyi identities}
\vskip 0.3cm

    For the special value $\kappa = 
    {1 \over \sqrt 2}$,  the equations 
   for $\psi$ and ${\vec A}$  can be reduced to  first 
   order differential equations. This special point was first
  used   by Sarma \cite{sarma} in his  discussion 
   of type-I vs type-II superconductors and then identified by
 Bogomol'nyi \cite{bogo} in 
  the more general context of stability and integrability of classical
 solutions of some quantum field 
  theories. This special point is also called a duality
 point.  We first  review  some  properties of
 the Ginzburg-Landau free energy at  the duality point. We 
   use   the following identity true 
   for two dimensional systems
\begin{equation} 
|({\vec{\nabla}} - i{\vec A})\psi {|^2}= |{\cal D} \psi {|^2} +
 {\vec \nabla} \times {\vec \jmath} + B |\psi{|^2}
\label{ident}
\end{equation}
where ${\vec \jmath}= Im( {{\psi^*}} {\vec \nabla} \psi )
   - |\psi{|^2}{\vec A}$ 
is the current density and the operator $\cal D$ is 
defined as  ${\cal D} = \partial_x + i\partial_y -i(A_x + iA_y)$. 
  This relation is a relative 
  of the Weitzenb\"ock formula (\ref{W}).
  At the duality point  
$\kappa = {1 \over \sqrt 2}$ the
 expression (\ref{gladim}) for   $\cal F$  can be rewritten using (\ref{ident}) as
\begin{equation}
{\cal F} = {\int_{\Omega}}
 \left(  {1 \over 2} | B - 1 + |\psi{|^2}{|^2} +
  |{\cal D} \psi {|^2} \right) \,\, 
 + {\oint_{\partial \Omega}} ({\vec\jmath} + {\vec A}).{\vec dl}
 \label{identitebog}
\end{equation}
where the last integral over the boundary ${\partial \Omega}$ of the
 system results from  Stokes theorem. 
 For an infinite system, we impose \cite{bogo} the usual 
 conditions for a superconductor, 
 namely $|\psi| \to 1$ and
 ${\vec \jmath} \to 0$  at infinity. 
  The boundary term in (\ref{identitebog}) then becomes
  \begin{equation}
  {\oint_{\partial \Omega}} ({\vec \jmath} + {\vec A}).{\vec dl} =
  {\oint_{\partial \Omega}} 
  ( { {\vec\jmath} \over  {|\psi|^2}} 
  + {\vec A}).{\vec dl} \
 \label{fluxoid}
 \end{equation}
  This last integral is the fluxoid. It is   quantized and is 
    equal to
 $$\oint_{\partial\Omega} {\vec \nabla}\chi.{\vec dl}= 2 \pi n \,\,\, . $$
  The integer $n$ is 
   the winding number of the order parameter $\psi$ and as such is a
 topological characteristic of the system.  Using (\ref{fluxoid}),
 we  see that $n$ is also the total magnetic 
 flux through the system:
\begin{equation}
  {\int_{\Omega}} B = n \,\,\, .
 \label{fluxquant}
 \end{equation}
  As we interpreted $B$ as a first Chern class, this relation
 is similar to  (\ref{chernc}).
   The   extremal values of ${\cal F}$, 
  are  obtained when the two
    Bogomol'nyi \cite{bogo}
  equations  are  satisfied
\begin{eqnarray}
 {\cal D} \psi & = &  0 
   \nonumber  \\
B & = &  1 - |\psi{|^2}.
\label{equaBogo}
\end{eqnarray}   
 In this case, the free energy  ${\cal F}$ is given by
 \begin{equation}
 \frac{1}{ 2 \pi } {\cal F} =  n 
 \label{freeenergie}
 \end{equation}
 Therefore the free energy itself is a topological invariant.
 The two Bogomoln'yi 
 equations can be decoupled
 and   $|\psi|$ is  a  solution
 of the second order
  nonlinear equation 
\begin{equation}
{\nabla^2} ln |\psi{|^2} = 2 ( |\psi{|^2} -1)
\label{Liouville}
\end{equation}
which is related to  the Liouville equation.

It should be noticed that the  set of equations 
(\ref{equaBogo},\ref{Liouville})
  has been  obtained without any assumption on the
 nature of the magnetic field and  appears
 in various other situations,  e.g. Higgs \cite{taubes},
 Yang-Mills \cite{witten} and
 Chern-Simons \cite{jackiw} field theories.  It 
was proven that these equations  admit families of vortex
 solutions \cite{taubes}. 
 For infinite systems,
 it can be shown  that each vortex carries one flux quantum
 and that the winding number  
 $n$ is equal to  the number of vortices in the system.
 However  for an infinite system
 there is no mechanism to  select the value of $n$.
 It will be  precisely the role of the 
 boundary of a finite system to introduce such a 
 selection mechanism
 and to determine $n$, according to the applied
 magnetic field.
   
\vskip 0.3cm

 \section {The superconducting billiard}
\vskip 0.3cm

  From  now on, we shall  study finite size
  systems in an external magnetic field $B_e$ and we denote by 
$\phi_e = \pi R^2 B_e / \phi_0 $ its dimensionless 
 magnetic flux through the system.
  The question then arises to know if they
  can sustain stable vortex solutions and 
  what   their behaviour is, as a function of the applied field.
  An interesting approach has been developed 
 by Bethuel and coworkers \cite{bethuel}.
They considered the case of a billiard $\Omega$
 without applied magnetic field but with the
 boundary 
condition for the order parameter $\psi |_{\Omega} = g (\theta) $ 
   with $g (\theta) =  e^{i \phi (\theta)}$ and 
 a prescribed 
 winding number ${1 \over {2i \pi}} \int_{0}^{2 \pi}
{ {\partial g} \over {\partial \theta}} d \theta = n \,\, . $
In the London limit, i.e. 
$\kappa \to \infty$, $|\psi|$ is 1 almost
 everywhere but because of the degree $n$
 on the boundary, $|\psi|$  must vanish $n$ times in the
  bulk therefore leading to vortices. 
An extension of this approach to the case where a magnetic field
 is applied on the system 
has been proposed in  \cite{serfati} where it is  shown  by
 a variational argument that vortex solutions 
 have a lower energy when the magnetic field is increased.
 By the same method, it is also 
possible to discuss the type of vortices
 and their distribution as a function of the geometry 
   of the billiard. Numerical simulations \cite{argentins}
 of the Ginzburg-Landau equations 
  for a long parallelepiped   in a uniform magnetic field show that
 the physical picture  derived for
   $\kappa = {1 \over \sqrt 2}$, namely
 the existence of stationary vortex solutions 
  whose number depends on the magnetic field,
  remains valid  for quite a large range of values of  $\kappa$,
 and the corresponding  change of free  energy is  small \cite{rebbi}. 
 We shall therefore study the case 
 $\kappa = {1 \over \sqrt 2}$, i.e. the duality
 point and extend the previous approach to a
  system with finite size where
  boundary effects are important.

 \subsection{The zero current line}

 In  a finite system, there are in general
 non-zero  edge currents
 and the order parameter is not equal to
 1 on the boundary. Hence, the identification of the boundary
 integral in (\ref{identitebog}) with the fluxoid (\ref{fluxoid})
 is not possible anymore, and the 
  free energy can not be minimized
 just by imposing Bogomol'nyi equations (\ref{equaBogo}).
 However, the  currents
 on the boundary of the  system screen the external magnetic field and 
  therefore produce a magnetic moment (a circulation)  
  opposite to the direction of the field, 
 whereas  vortices in the bulk of the system produce a
 magnetic  moment along the direction of 
  the applied field.
 Hence  currents in the bulk  circulate in a direction opposite
 to those  at the boundary.
 If one assumes cylindrical symmetry,  $\vec\jmath$   has only 
 an azimuthal component, with  opposite  signs  in the bulk
 and on the edge of  the system 
 (the  radial component is  zero
 since  ${\vec\jmath}$ is  divergence free).
 Thus,  there exists  a circle $\Gamma$
  on which  $\vec\jmath$ vanishes.
 This allows us to separate  the domain $\Omega$
  into two concentric subdomains
 $\Omega = {\Omega_1}
\cup {\Omega_2}$ 
 such that the boundary  $\partial \Omega_1$  is  the curve
 $\Gamma$.  On $\partial \Omega_1$, the current density
  ${\vec \jmath}$ is zero, therefore  
\begin{equation}
{\oint_{\partial \Omega_1}} {\vec \jmath}.{\vec dl}  =
  {\oint_{\partial \Omega_1}} 
   { {\vec \jmath} \over  {|\psi|^2}}.{\vec dl} = 0 \\ .
 \label{petitbogo}
\end{equation}

 Thus  one deduces as above
  that Bogomol'nyi and Liouville equations
  are valid in the finite domain
 $\Omega_1$ as in the case of the infinite plane.
The magnetic flux $\Phi ({\Omega_1})$  is calculated
 using the fluxoid  and (\ref{petitbogo}) so that  
 $$\Phi ({\Omega_1}) =  
 n - \oint_{\partial{\Omega_1}} {{{\vec \jmath}.{\vec dl}}
 \over {|\psi{|^2}}} =   n .$$
 As before  $n$ 
  is  the 
winding number, i.e. 
 $\oint_{\partial{\Omega_1}} {\vec \nabla }\chi.{\vec dl} =
 2\pi n$, as well as  the number of vortices \cite{am} in $\Omega_1$.   
 The free energy in $\Omega_1$  is 
\begin{equation}
{\cal F}({\Omega_1})  = 2 \pi n .
\label{free1}
\end{equation}

The contribution of  $\Omega_2$ to the free energy 
is given by  (2) and can be expressed using
the phase and the modulus of the order parameter $\psi$ 
 \begin{equation}
{\cal F}(\Omega_2)={\int_{\Omega_2}}  (\nabla |\psi|{)^2}
 + |\psi{|^2} |{\vec{\nabla}}\chi - {\vec A}{|^2}
 + {{B^2} \over 2} + { {(1 - |\psi{|^2}{)^2}} \over 2}
\label{tran1}
\end{equation}

   The boundary conditions for both the magnetic field $ B(R)$ 
 and the vector potential $A(R)$  adapted 
  to a flat disk geometry are provided by
 the condition $\phi = {\phi_e}$. It implies
  that at the boundary $B(R)$ is larger than the applied field $B_e$ due 
  to the distorsion of the flux lines near the edge of the system.

 \subsection{A selection mechanism and topological phase transitions}

The analysis presented in \cite{am} leads for
 the total Gibbs potential of 
the billiard to the expression
\begin{eqnarray}
{1 \over {2 \pi}} {\cal G}(n, {\phi_e}) &=&
 {1 \over {2 \pi}} {\cal F}(n, {\phi_e}) -  {{ 2 \lambda^2}
 \over {R^2}}{\phi_e}^2 \nonumber \\
  &=&   n + {{ \lambda {\sqrt 2}} \over R} (n - {\phi_e} {)^2}
 - {1 \over 2} ({{\lambda {\sqrt 2}} \over R}{)^3}(n - {\phi_e} {)^4}
  - {{ 2 \lambda^2}
 \over {R^2}}{\phi_e}^2
\label{gibbs}
\end{eqnarray}
This relation consists in a set of quartic
 functions indexed by the integer $n$. The minimum 
of the Gibbs potential is the envelop curve defined by the equation 
${{\partial {\cal G}} \over {\partial n}}{|_{\phi_e}} =0$ (using obvious notations),
 i.e. the system chooses its winding number 
$n$ in order to minimize $\cal G$. This provides a relation between 
the number $n$ of vortices in the system and the
 applied magnetic field $\phi_e$. 
In the limit of a large enough $ R \over \lambda$,
 the quartic term is negligible and the 
Gibbs potential reduces to a set of parabolas.
 The winding number $n$ is then given by
\begin{equation}
    n = [  { \phi_e}
 - {R \over {2 {\sqrt2}\lambda}} + {1 \over 2}  ] 
\end{equation}
 The magnetization of the  
system 
   $ M= - {  {\partial {\cal G}} \over  {\partial \phi_e}  }$  
 is given by
\begin{equation}
- M = {{2{\sqrt 2} \lambda} \over R} ( {\phi_e} - n ) -
 {{ 4 \lambda^2} \over {R^2}} {\phi_e}
\label{magnet}
\end{equation}
For ${\phi_e}$ smaller that
  $R \over {2{{\sqrt 2}\lambda}}$, we have $n =0$
 and  $(- M)$  increases linearly with 
 the external flux. This corresponds 
to the London regime before the first vortex enters the system.
 The field $B_1$ at which the first vortex enters the system 
 corresponds to ${\cal G}(n=0)=
 {\cal G}(n=1)$, i.e. to 
${B_1} = { {\phi_0} \over {2 \pi {\sqrt 2} R \lambda} }
 + { {\phi_0} \over {2 \pi  R^2} }$.
 The subsequent 
vortices enter one by one for each crossing
 ${\cal G}(n+1)= {\cal G}(n)$; 
 this happens periodically in the applied field,
 with a period equal to
 ${\Delta}H = {{\phi_0}
 \over { \pi  {R^2}}}$. This gives rise to 
a discontinuity of the magnetization
 $ {\Delta}M = {{2 {\sqrt 2} \lambda} \over R }.$

\subsection{A geometrical  expression of the Gibbs
 potential for finite systems}

  For  infinite system, with the boundary conditions
 $|\psi| \to 1$ and ${\vec \jmath} \to 0$  at infinity, 
 we showed that  the free energy at the dual point 
  $\kappa = {1 \over {\sqrt 2}}$
 is  a topological invariant 
 proportional (\ref{freeenergie})   to the fluxoid $n$, which
    represents also the number of vortices in the system:
\begin{equation}
  \frac{1}{ 2 \pi } {\cal F} = \int B =   n \,\,\, .
\label{energ1}
 \end{equation}
  This relation  is a property of the dual point
 at which the Ginzburg-Landau functional 
has a geometrical interpretation and admits
 the quantized magnetic flux as a topological invariant.
 For  a finite system the fluxoid quantification 
 can be expressed as
 \begin{equation}
{1 \over {2 \pi}} {\int_\Omega} {\vec B}.d{\vec S} +
\oint_{\partial \Omega} {{\vec  \jmath} \over {|\psi{|^2}}}.d{\vec l}
   = n  \,\,\, .
\end{equation}
This relation  is analogous to the Gauss-Bonnet theorem for
a surface  with a boundary  (\ref{gb}) or more generally
 to a  topological invariant obtained by
 summing  a Chern class in the bulk and a Chern-Simons
 class on the boundary (\ref{simon}). Recalling
 (\ref{fluxo1}), we see that 
 the magnetic field $B$  plays the role of a  curvature $K$, and
 the quantity ${{\vec  \jmath} \over {|\psi{|^2}}} = \nabla\chi - A$
 is similar to a  {\it geodesic curvature} $k_g$.
 For a system with cylindrical symmetry, one can show \cite{am}
  that  ${{\vec  \jmath} \over {|\psi{|^2}}} = n - {\phi_e}$.
 Therefore equation  (\ref{gibbs}) can be rewritten as
\begin{equation} 
{1 \over {2 \pi}} {\cal F} = \int_{\Omega}  B   + \int_{\partial\Omega}
\eta( {  {\vec  \jmath} \over {|\psi{|^2}}} ) 
  \equiv  \int K + \oint \eta(k_g)   \,\,\, .
\label{energiegeom} 
\end{equation}
 Comparing (\ref{energiegeom}) with (\ref{energ1})
 we conclude that 
 the boundary correction is a  functional  
 of the geodesic 
  curvature. In the preceeding section, 
 we  obtained  an  explicit formula   for 
 the  function $\eta$ as an  even  fourth order polynomial in 
 the geodesic curvature.

 This geometric interpretation makes us believe that an expression such as
 (\ref{energiegeom}) is fairly general. It could be well suited,
 as an Ansatz, to describe finite systems which are known to have
 a topological description in the infinite limit; for example,
 a suitable  generalization of (\ref{energiegeom}) to SU(2) symmetry
 could  describe superfluid $^3$Helium in a bounded domain.

We have presented a geometrical formulation for the  
Ginzburg-Landau problem which we now briefly summarize.

i) For a certain theory, like Ginzburg-Landau 
or other functionals of this type, it may 
appear stable singular solutions (e.g. vortices)
 whose nature is determined  by the related homotopy groups of the Toulouse-Kleman \cite{toulouse} approach.
To these solutions are associated topological numbers
 in the Bogomol'nyi limit.

ii) The existence of  topological numbers signals
 the occurence of a geometrical description of the problem.

What is it good for? First, we notice that topological 
quantities describe global features 
of the problem i.e. behaviour in the large by opposition
 to the local behaviour 
obtained from solutions of differential equations.
 Then, by identifying physical 
quantities in terms of global topological invariants,
 we do not need to solve the local 
equations to obtain the behaviour of the system.
As a result, we may say that an important goal of 
a geometrical description  
is to obtain physical quantities in terms of global topological 
expressions. When it is possible, it is very much rewarding.
 
{\bf Acknowledgments :}

It is our pleasure to thank Alain Joets for his critical reading of the manuscript.

\eject


\begin{thebibliography}{article}
\bibitem{remarque} Generally, $d$ and $\delta$ are only 
 formally adjoint. The integration by parts gives
  is $ (\phi_k, d \psi_{k-1}) - (\delta \phi_k, \psi_{k-1})
   = \int_{M} d ( \* \phi_k \wedge \psi_{k-1})$. 
  The rhs integral is a boundary term and vanishes for a compact
    manifold $M$ without boundary; in that case  $\delta$ and 
   $d$ are truly adjoint.
\end{thebibliography}
\end{document}